\documentclass[onecolumn]{revtex4}
\usepackage{ragged2e}
\usepackage{amsmath}
\usepackage{graphicx}
\usepackage{bm}

\usepackage[usenames,dvipsnames]{color}

\definecolor{darkblue}{RGB}{0,0,196} 
\usepackage[colorlinks=true,linkcolor=darkblue,citecolor=darkblue,urlcolor=darkblue]{hyperref}

\begin{document}

\title{Two Key Day-One Observables in Electron-Nucleus Collisions at the Upcoming Electron-Ion Collider and Their Implications for Baryon Number Carrier Investigations}\vspace{0.5cm}

\author{Ting-Ting~Duan$^{1,}${\footnote{202312602001@email.sxu.edu.cn. 
ORCID ID 0009-0005-4353-0655}},
Fu-Hu~Liu$^{1,}${\footnote{Correspondence: fuhuliu@163.com;
fuhuliu@sxu.edu.cn. ORCID ID 0000-0002-2261-6899}},
Khusniddin~K.~Olimov$^{2,3,}${\footnote{Correspondence:
khkolimov@gmail.com; kh.olimov@uzsci.net. ORCID ID
0000-0002-1879-8458}} }

\affiliation{$^1$Institute of Theoretical Physics, State Key
Laboratory of Quantum Optics Technologies and Devices \&
Collaborative Innovation Center of Extreme Optics, Shanxi
University, Taiyuan 030006, China
\\
$^2$Laboratory of High Energy Physics, Physical-Technical
Institute of Uzbekistan Academy of Sciences, Chingiz Aytmatov Str.
2b, Tashkent 100084, Uzbekistan
\\
$^3$Department of Natural Sciences, National University of Science
and Technology MISIS (NUST MISIS), Almalyk Branch, Almalyk 110105,
Uzbekistan
}

\begin{abstract}

\vspace{0.5cm}

\noindent {\bf Abstract:} 
In high-energy collisions, the transverse momentum and rapidity (or pseudorapidity) distributions of charged particles are two key day-one observables related to characterize the collective dynamics of the produced system, and the upcoming Electron-Ion Collider (EIC) is no exception. While these distributions are sensitive to collective behavior, previous studies have demonstrated that they also exhibit distinct features at different energies that are intrinsically correlated with the nature of baryon number carriers: whether they are valence quarks or gluon junctions. Across the energy range from a few GeV to tens of GeV, extending to hundreds of GeV and even the TeV scale, baryon number carriers may undergo a partial or complete transition from valence quarks to gluon junctions. Understanding this transition is essential for a comprehensive picture of particle production, which in turn provides a crucial baseline for disentangling genuine collective effects in future EIC experiments.
\\
\\
\end{abstract}

\maketitle

\parindent=15pt

\section{Introduction}

High-energy collisions represent a significant area of research in modern physics~\cite{1,2,2a,2b}, allowing for the investigation of bulk properties of multiple particles through various theoretical models~\cite{3,3a} and technical methods~\cite{3b,4,5}. These bulk properties encompass a range of characteristics including, but not limited to, multiplicity distributions, invariant yields or transverse momentum distributions, rapidity and pseudorapidity distributions. As key observables, these distributions not only provide essential insights into particle production mechanisms but also serve as crucial probes for characterizing the collective dynamics of the produced system (such as the dependence of anisotropic flow on transverse momentum). The models employed to describe these phenomena can be categorized into several types such as transport and hydrodynamic models, relativistic and quantum molecular dynamics (QMD) models, along with thermal and statistical models. To derive numerical results regarding the evolution characteristics of collision systems and the distribution laws governing multiple particles, these related models are often implemented using Monte Carlo methods.

In the theoretical modelling analysis of high-energy collisions, certain nuclear structures, alongside nucleon structures~\cite{6,7,8}, can play very important roles~\cite{9,10}. Nuclear structures include factors such as \(\alpha\)-clustering within nuclei, non-uniform number densities of nucleons, as well as shapes and orientations associated with deformed nuclei. Nucleon structures comprise aspects like types of baryon number carriers, spin and magnetic moments associated with nucleons and their constituents, in addition to current masses and constituent masses attributed to quarks. Notably, different carriers for baryon number may lead to variations in multiplicity, transverse momentum, and (pseudo)rapidity distributions, due to differing penetrability levels exhibited by projectiles or stopping power experienced by targets.

There exist two potential carriers for baryon number: valence quarks~\cite{11,11a,11b,11c} and the gluon (or baryon) junction~\cite{12,12a,13,13a,13b,14,15,16,16aa,16a}, although neither has been conclusively verified thus far~\cite{17,18,18a}. Within the standard framework provided by quantum chromodynamics (QCD), each valence quark is understood to carry one-third of the total baryon number~\cite{11,11a,11b,11c}, which form a triangular configuration known as the \(\triangle\)-shaped topology. Each valence quark is positioned at one tip of the triangular topology, with a Wilson line connection established between each pair of valence quarks. An alternative proposal suggests that baryon number may be carried by a non-perturbative configuration of gluon fields, referred to as the gluon~\cite{12,12a,13,13a} or baryon junction~\cite{13b,14,15,16,16aa,16a}. This structure is assumed to be gauge-invariant and located at the center of the Y-shaped topology. In this scenario, each valence quark resides at one tip of the Y-shaped topology, and there exists a Wilson line connection between each valence quark and the gluon junction.

In our view, irrespective of whether baryon number is carried by valence quarks and/or gluon junctions, for transverse momentum (\(p_T\)) from low (\(p_T < 2\) GeV/\(c\)) to high (\(p_T > 8\) GeV/\(c\)), for rapidity from the central region to the fragmentation (backward/forward) region, and for collision energy from a few GeV to above TeV, the carriers may change in different local transverse momentum, rapidity, and/or energy ranges. We should pay special attention to the transverse momentum and rapidity distribution ranges of charged particles produced in collisions at a given energy when studying the information of baryon number carriers. Based on our perspective, the seemingly contradictory literature on different types of baryon number carriers is actually not contradictory, as their focused transverse momentum and rapidity ranges may not be consistent.

In the region \(p_T < 2\) GeV/\(c\), particle production is mainly dominated by soft non-perturbative processes, and the distribution presents an approximately exponential Boltzmann-type spectrum. This is governed by the thermal freeze-out mechanism of the hadronization process, corresponding to multiple soft parton interactions. This is a typical feature of the low-energy range where valence quarks act as the dominant baryon number carriers. In the intermediate region (\(2 < p_T < 8\) GeV/\(c\)), particle production involves contributions from both soft processes and perturbative hard scattering. The distribution gradually transitions from an exponential form to a power-law form. It is worth noting that in this region, significant collective flow effects (such as elliptic flow \(v_2\)) can be observed, concurrently with a significant enhancement of baryon yield relative to mesons. The latter is directly related to the participation of gluon junctions as baryon number carriers in these processes. Disentangling the signals of baryon number carriers from the background of collective dynamics in this \(p_T\) range is a challenging but crucial task. In the region \(p_T > 8\) GeV/\(c\), particle production is dominated by the perturbative QCD hard scattering process, and the distribution follows a power-law decreasing QCD scaling behavior, corresponding to the hard scattering process of valence quarks and hard gluons in the large Bjorken \(x\)-range. The baryon number carriers are valence quarks in the large \(x\)-range, and the jet quenching effect is significant in nucleus-nucleus collisions.

In the energy range of several GeV, the rapidity distribution exhibits a clear bimodal structure, with two peaks located at the initial rapidity positions of the projectile and target nuclei, respectively. The particle yield in the central rapidity region is low, and the baryon number is mainly concentrated in the fragmentation region. Valence quarks, as baryon number carriers, are absolutely dominant. In the energy range of tens to hundreds of GeV, the double peaks gradually approach the central rapidity region, and the particle yield in the central rapidity region significantly increases. The distribution transitions from a bimodal structure to an approximately flat-topped platform structure, and the contributions of sea quarks and gluons begin to become prominent. Some baryon number carriers begin to transition from valence quarks to gluon junctions. In the energy range from several hundred GeV to several TeV and beyond, the rapidity distribution presents a broad, approximately Gaussian flat-top structure, with the highest particle yield and uniform distribution in the central rapidity region. The bimodal characteristics of the fragmentation region are significantly weakened, and the gluon-dominated parton evolution process completely takes the leading role. The contribution of gluons (via gluon junctions) as baryon number carriers is significantly increased.

In particular, in the energy range from a few GeV to tens of GeV, it is known that valence quarks will mainly manifest in the forward and backward rapidity regions due to their strong penetrability, when they act as non-principal participants in nucleon-nucleon (\(NN\)) collisions, where sea quarks and gluons are the principal participants. This type of collision is characterized as a soft excitation process, where the transverse momentum transfer is low (\(p_T < 1\) GeV/\(c\)). Comparatively, valence quarks are the absolute dominant participants in major hard scattering processes, where the transverse momentum transfer is high (\(p_T > 1\) GeV/\(c\)). Here, the principal and non-principal participant framework for partons in \(NN\) collisions is similar to the participant-spectator framework used in nucleus-nucleus (\(AA\)) collisions. At the nucleonic level in \(AA\) collisions, this participant-spectator picture~\cite{19,20,21} has been extensively utilized~\cite{22,23,24} for many years~\cite{25,26,27}.

Despite extensive research on charged particles, there is relatively little research on net-baryons. It can be expected that the distribution characteristics of net-baryons are similar to those of charged particles, but with the contribution of nucleons pre-existing in the projectile and target nuclei. From the perspective of collective dynamics, net-baryons (as the difference between baryons and anti-baryons) can effectively cancel out a significant portion of the background effects arising from soft, thermalized collective expansion, thereby providing a cleaner baseline. We are curious about the distribution characteristics of net-baryon transverse momentum and rapidity, which contain purer carrier information of baryon numbers.

In this review article, we analyze net-proton production in \(AA\) collisions at GeV energies and provide a prediction for net-proton production at the upcoming Electron-Ion Collider (EIC). The analysis is conducted using both soft and hard components within the framework of a multi-source thermal model~\cite{28,29}. Furthermore, we discuss potential carriers of baryon number with optimism that future investigations at the EIC will offer further validation. Finally, we summarize this work.

\section{Soft and hard components of particle distribution}

The multi-source thermal model~\cite{28,29} is one of the thermal
and statistical models, which is also a hybrid model using
distinct pictures and distributions for various charged particles
and nuclear fragments. In the model, multiple
principal participating or contributing quarks
and gluons can be regarded as the multiple energy sources at the
level of parton. In high-energy nuclear collisions, the basic
contributors in the nucleus are nucleons. Meanwhile, in the
nucleon or other hadrons, the basic contributors are partons. In
collisions induced by a lepton, the lepton is also a contributor
which is approximately equivalent to a parton.

Each or the $i$-th contributor energy source contributes a
quantity $p_{ti}$ to transverse momentum $p_T$ of charged
particles. Let $p_{ti}$ obey an exponential distribution, one has
\begin{align}
\label{eq.1} f_{p_{ti}}(p_{ti})=\frac{1}{\langle p_{ti}\rangle}
\exp\left(-\frac{p_{ti}}{\langle p_{ti}\rangle}\right).
\end{align}
In Equation~\ref{eq.1}, $\langle p_{ti}\rangle$ is the average of $p_{ti}$, which
results in the exponential distribution to be normalized to 1. A
subscript $p_{ti}$ is used in $f_{p_{ti}}(p_{ti})$ to distinguish
the distribution from others which will be discussed later.

If $p_{T}$ is contributed by $m_j$ contributors, the distribution
of $p_{T}$ is the fold of $m_j$ exponential distributions. One has
$p_{T}$ distribution to be an Erlang distribution, that
is~\cite{28}
\begin{align}
\label{eq.2} f_{p_T,E}(p_T)=\frac{p_T^{m_j-1}}{(m_j-1)!\langle
p_{tij}\rangle^{m_j}} \exp\left(-\frac{p_T}{\langle
p_{tij}\rangle} \right).
\end{align}
In Equation~\ref{eq.2}, $\langle p_{ti1}\rangle$ is for the first process, i.e., the soft excitation process, while $\langle p_{ti2}\rangle$ is for the
second process, i.e., the hard scattering process. Usually,
$\langle p_{ti1}\rangle$ is considered the smaller one in $\langle
p_{tij}\rangle$ ($j=1$ and 2). There is no limitation for the
relative size of $m_1$ and $m_2$, though $m_1\langle
p_{ti1}\rangle <m_2\langle p_{ti2}\rangle$. 

Over the energy range from several GeV to several hundred GeV, there are 
few ($m_1$) contributors (sea quarks, gluons and lepton) involved in the 
soft excitation process, and another few ($m_2$) contributors
(valence quarks and lepton) involved in the hard
scattering process~\cite{28}, where the lepton is included in $m_{1,2}$ if
it induces the collisions. One has a superposition of two Erlang
distributions to be
\begin{align}
\label{eq.3} f_{p_{T},2E}(p_{T})=\sum_{j=1,2} \frac{k_j
p_{T}^{m_j-1}}{(m_j-1)!\langle p_{tij}\rangle^{m_j}}
\exp\left(-\frac{p_{T}}{\langle p_{tij}\rangle} \right).
\end{align}
Here, $k_1$ ($k_2$) is the contribution fraction of the soft
excitation (hard scattering) process and $\sum_{j=1,2} k_j=1$. The
contribution of the first component distributes in a narrow region
around the low $p_T$, and the contribution of the second component
distributes in a wide region from the low to high $p_T$.

Based on Equation~\ref{eq.3}, one can obtain the average transverse momentum contributed by the soft process to be $\langle p_T\rangle_1=m_1\langle p_{ti1}\rangle$, and the average transverse momentum  contributed by the hard process to be $\langle p_T\rangle_2=m_2\langle p_{ti2}\rangle$. Obviously, $\langle p_T\rangle_1 <\langle p_T\rangle_2$. For a set of data sample, the average transverse momentum $\langle p_T\rangle$ contributed by the soft and hard processes is $\langle p_T\rangle =k_1\langle p_T\rangle_1+(1-k_1)\langle p_T\rangle_2= k_1 m_1\langle p_{ti1}\rangle +(1-k_1)m_2\langle p_{ti2}\rangle$. The fitted $\langle p_T\rangle$
reflects the overall effects of the thermal motion of the produced particles and the collective expansion of the collision system. In general, $\langle p_T\rangle$ is proportional to the temperature of the emission source of the produced particles and can be used to characterize the collision process.
The corresponding temperature parameter $T_1$ extracted from the soft process should be less than the temperature parameter $T_2$ extracted from the hard process. The average temperature extracted from the two process is $T=k_1T_1+(1-k_1)T_2$.  

At least two contributors taking part in the collisions, both the
minimum values of $m_1$ and $m_2$ are 2. In some cases, $k_1=1$,
which means that there is no contribution of the second component.
If $k_1<1$, one has to consider the contribution of the second
component. Although the maximum value of $m_1$ is not limited,
this value is comparable to $m_2$ according to our
investigation~\cite{28,29}. The maximum value of $m_2$ is
4 or greater if 4 or more valence quarks in the
projectile and target nucleons take part
principally in the collisions in the considered energy range, though the
probability of this occurring is very low. Because
the collision energy discussed in this work is not too high (less than 
several hundred GeV), the most likely scenario is that two valence quarks 
in two nucleons take part principally in the collisions,
resulting in $m_2=2$.

Although $p_T$ distribution of charged particles can be fitted by
few functions~\cite{30,31}, the two-component Erlang distribution
in the framework of multi-source thermal model~\cite{28,29} can
also be used. To see variable shapes of curves from Erlang
distribution and its two-component form, our previous work studied
the examples with different parameters in related distribution
with both the linear and logarithmic coordinates~\cite{28}, as
well as respective contributions of the first and second
components and their superposition in $p_T$
distribution~\cite{29}. At the EIC, there is no particular change
in the shape of experimental $p_T$ spectra of charged particles.
To avoid unnecessary repetition if specific parameter values are
not available, no relevant curves are provided here. From our
previous work~\cite{28,29}, one can see the abundant results
related to Erlang distribution. Indeed, the two-component Erlang
distribution is very flexible in the fit to $p_T$ distributions.
Similar discussions and formulas also apply to multiplicity
distributions.

We would like to clarify that in the two-component model for $p_T$
distribution (Equation~\ref{eq.3}), the normalization related to
system volume is not two values, despite the two components
corresponding to different $p_T$ regions. Specifically,
Equation~\ref{eq.3} serves as a probability density function, where
$f_{p_{T},2E}(p_{T})$ can be used to compare with experimental
$p_T$ spectra and determine a single normalization value. Even if
there are two system volumes $V_1$ and $V_2$, they relate to two types of events:
one ($V_1$) from soft excitation process and another ($V_2$) from hard scattering
process. The total volume or normalization constant is that $V=V_1+V_2$. 
Moreover, it is noted that experimental spectra exhibit a
rapid or exponential decrease in intermediate- and high-$p_T$
regions. This results in the fraction of soft excitation processes significantly 
exceeding that of hard scattering processes. Including
data at higher $p_T$, the dominance of soft excitation process
remains due to a very small fraction in the extremely high-$p_T$
region.

On the rapidity ($y$) or pseudorapidity ($\eta$) distribution of
charged particles, in collisions in our considered energy range, 
the soft excitation process which involved to
sea quarks and gluons leads to a wide range from the backward to
forward rapidity regions due to the penetrability of the
non-principal participating valence quarks.
Correspondingly, the hard scattering process which involved to
the principal participating valence quarks leads to
a higher probability in the central rapidity region,
though other partons results in a small amount of charged particles 
distributed in the whole rapidity region. In the considered collisions,
baryons have higher probability appearing in the backward and
forward rapidity regions due to the contribution of leading
nucleon effect, which is due to more events led by soft excitation
process.

Generally, experimental data measured by international
collaborations are a mixture of the soft excitation and hard
scattering processes. From the backward rapidity region to the
central one, then to the forward one, charged particles distribute
in a wide range. In the rest frame of the emission source,
particles are assumed to be emitted isotropically. According to
the 1+1-dimensional hydrodynamic model firstly proposed by
Landau~\cite{31a}, the rapidity distribution of charged particles
produced in the emission source with rapidity $y_x$ obeys a
Gaussian form~\cite{31b},
\begin{align}
\label{eq.4} f_{y,G}(y)=\frac{1}{\sqrt{2\pi}\sigma_x}
\exp\left[-\frac{(y-y_x)^2}{2\sigma_x^2} \right],
\end{align}
where $\sigma_x$ is the distribution width or standard deviation,
and $y_x=0$ corresponds to the rest frame of the
emission source. Equation~\ref{eq.4} is normalized to unity when integrated from $-\infty$ to $+\infty$. 

Let $y_T$, $y_C$ and $y_P$ be the rapidities of emission sources
located at the backward (target), central and forward (projectile)
rapidity regions, respectively. The rapidity distribution measured
in final state is the sum of three Gaussian distributions. That is
\begin{align}
\label{eq.5} f_{y,3G}(y)=\frac{1}{\sqrt{2\pi}}\sum_{x=T,C,P}
\frac{k_x}{\sigma_x} \exp\left[-\frac{(y-y_x)^2}{2\sigma_x^2}
\right],
\end{align}
where $k_{T,C,P}$ are the contribution fractions of the emission
sources with $y_{T,C,P}$, and $\sum_{x=T,C,P}k_x=1$. The emission
source with $y_C$ is related to both the soft
excitation and hard scattering processes.
Specifically, the soft process contributes
significantly to low-$p_T$ particles, while the hard process makes
a considerable contribution to high-$p_T$ particles. The emission
sources with $y_{T,P}$ are dominated by the soft process,
with the hard process contributing very little.
Due to large $\sigma_x$, the contributions of three sources can be
overlapped, and at least the contributions of two adjacent sources
can be overlapped. As one of the most common distributions, we
have applied the superposition of Gaussian distributions, Equation~\ref{eq.5}, in our
previous work~\cite{31c,31d}. When treating the
three emission sources as a whole, a wider Gaussian distribution
is obtained, with a width approximately equal to
$\sqrt{\ln(\sqrt{s_{NN}}/2m_p)}$~\cite{31a,31e}, where
$\sqrt{s_{NN}}$ denotes the center-of-mass energy per nucleon pair
and $m_p$ is the rest mass of the proton.

It should be noted that, a unified Erlang distribution is used to
describe both the soft and hard components of $p_T$ distribution
of charged particles produced in high-energy collisions. The total
result is the superposition of two Erlang distributions in which
the smaller (larger) $m_j\langle p_{tij}\rangle$ correspond to the
contribution of the first (second) component. Although the two
components correspond to different intensities of collisions, both
the contributors are partons (and lepton in electron induced
collisions at the EIC if available) which are regarded as the
energy sources of particle production. Meanwhile, a unified Gaussian $y$ distribution is used for particles in the backward, central and forward rapidity regions. The unified form is a reflection of the similarity, commonality and
universality~\cite{32,33,34,35} existed in high-energy
collisions~\cite{36,37,38,39}.

In the case of considering $eA$ collisions at the EIC, which is in
fact electron-nucleon ($eN$) or electron-proton ($ep$) and
electron-neutron ($en$) scattering, one expects that $m_1$ in
Equation~\ref{eq.3} will be 2--3, as the projectile $e$ and 1--2 sea
quarks or gluons from the target are likely to be
the principal participants in the collisions. Regarding the hard
scattering process, we expect $m_2=2$, which reflects the
contributions of both the projectile $e$ and the
principal participating valence quark from the target
in the collisions.

It is expected that $p_T$ distribution of charged particles
produced in $eA$ collisions will follow Equation~\ref{eq.3}. Our
previous work demonstrates that in the central
rapidity region, the soft component accounts for 60--70\% of the
yield at TeV energies~\cite{29}. It is anticipated that the
contribution fraction of the soft component will be even higher at
the EIC due to lower energy. Because the specific value of
$\langle p_{tij}\rangle$ is not yet clear, we could not provide a
specific curve here. However, one may refer to our previous work
to understand the trend of the curve~\cite{28,29}.

The rapidity distribution of charged particles produced in $eA$
collisions can also be described by Equation~\ref{eq.5}. However, an
asymmetric distribution will be observed, in which a small yield
appears in the forward rapidity region ($e$-going direction) due
to the projectile only including one
participating $e$, and a large yield occurs in
the backward rapidity region ($A$-going direction) due to the
target containing more principal participating
partons. This results in $k_P<k_T$ in Equation~\ref{eq.5}.

At the same time, the peak position of rapidity distribution will
be shifted to the backward rapidity region. In other words,
charged particles from the soft excitation process are mainly
distributed in the backward and central rapidity regions, and
those from the hard scattering process are mainly distributed in
the central rapidity region. Considering the larger average $p_T$ 
($\langle p_T\rangle$) of charged particles in the hard scattering
process, central rapidity region corresponds to larger $\langle
p_T\rangle$ than other rapidity regions, which results in higher
temperature of emission source in central rapidity region.

In particular, for net-baryons produced in $eA$ collisions, one
has $k_P=0$, Equation~\ref{eq.5} is then changed to a two-component
form. This means that net-baryons are only distributed in the
backward and central rapidity regions, but not in the froward
rapidity region, because there is no leading nucleon in the
projectile $e$. Comparing with the soft excitation process, the
larger $\langle p_T\rangle$ of baryons can be observed in the hard
scattering process, which results in higher temperature of
emission source.

It should be noted that the classification of soft excitation and
hard scattering processes is a broad categorization. In reality,
events can be subdivided into numerous groups based on an
increasing collision strength, ranging from the softest to the
hardest processes. We can consider an alternative approach which
contains more components. Particles generated in the softest
process are anticipated to manifest in both the most backward and
forward rapidity regions, characterized by a lower temperature of
their emission source. Conversely, particles produced in the
hardest process are expected to emerge predominantly
at mid-rapidity, where the temperature of their
emission source is at its highest. Most particles
are located near the mid-rapidity or within central rapidity
region, and their emission source temperature is high.

In the case of considering numerous groups of events, in which
each group forms a source with rapidity shift $y_x$ and
temperature $T_x$, one may use a multi-Erlang distribution for
$p_T$ spectra and a multi-Gaussian form for $y$ distribution.
Alternatively, each source or component can be also described by
an ideal gas model which is isotropic in the rest frame of the
source. One has the unity-density of $p_T$ and $y$ to be~\cite{30}
\begin{align}
\label{eq.6} \frac{d^2N}{dydp_T} =\frac{gV_x}{(2\pi)^2}p_T
\sqrt{p_T^2+m_0^2} \cosh (y-y_x)
\bigg\{\exp\bigg[\frac{\sqrt{p_T^2+m_0^2} \cosh
(y-y_x)-\mu}{T_x}\bigg]+S\bigg\}^{-1},
\end{align}
where $g=2s+1$ is the degeneracy factor, $s$ is
the spin quantum number (which is equal to 0 for $\pi^{\pm}$ and
$K^{\pm}$ mesons, and 1/2 for proton and neutron), $V_x$
represents the volume of the source, $m_0$ denotes the rest mass
of the considered particle, and $\mu$ is the chemical potential of
the given particle. In addition, $S=-1$, 1, and 0 correspond to
Bose-Einstein, Fermi-Dirac, and Maxwell-Boltzmann statistics,
respectively.

From the unity-density Equation~\ref{eq.6}, the density of $p_T$ can
be written as
\begin{align}
\label{eq.7} \frac{dN}{dp_T} =\frac{gV_x}{(2\pi)^2} p_T
\sqrt{p_T^2+m_0^2} \int_{y_{\min}}^{y_{\max}} \cosh (y-y_x)
\bigg\{\exp\bigg[\frac{\sqrt{p_T^2+m_0^2}\cosh
(y-y_x)-\mu}{T_x}\bigg]+S\bigg\}^{-1}dy,
\end{align}
where $y_{\min}$ and $y_{\max}$ are the minimum and maximum
values of $y$. Mathematically, $y_{\min}$ and $y_{\max}$ in Equation~\ref{eq.7} can be negative and positive infinite respectively. In fact, large enough $|y_{\min}|$ and $y_{\max}$ are usually used in the fit. Similarly, the density of $y$
from the unity-density is
\begin{align}
\label{eq.8} \frac{dN}{dy}
=\frac{gV_x}{(2\pi)^2}\int_0^{p_{T\max}} p_T \sqrt{p_T^2+m_0^2}
\cosh (y-y_x) \bigg\{\exp\bigg[\frac{\sqrt{p_T^2+m_0^2}\cosh
(y-y_x)-\mu}{T_x}\bigg]+S\bigg\}^{-1}dp_T,
\end{align}
where $p_{T\max}$ is the maximum $p_T$. Although $p_{T\max}$ in Equation~\ref{eq.8} can be infinite mathematically, a large enough $p_{T\max}$ is usually used in the fit.   

The probability density function of $y$ is given by
\begin{align}
\label{eq.9} f(y,y_x,T_x)=\frac{1}{N}\frac{dN}{dy}.
\end{align}
Equation~\ref{eq.9} is the normalized $y$ distribution of particles produced in
the source with $y_x$ and $T_x$. In the final $y$ distribution from the whole collision system which is distributed in a wide longitudinal range, the contribution fraction or weight of the source with $y_x$ and $T_x$ is proportional to $V_x$. For general charged particles such as $\pi^{\pm}$ and $K^{\pm}$, the weight are the same. Correspondingly, one has the final $y$ distribution to be
\begin{align}
\label{eq.10} f(y)=\frac{1}{N}\frac{dN}{dy}=\frac{k}{y^{\rm
Bwad}_{\max}-y^{\rm Bwad}_{\min}}\int_{y^{\rm
Bwad}_{\min}}^{y^{\rm Bwad}_{\max}} f(y,y_x,T_x)dy_x
+\frac{1-k}{y^{\rm Fwad}_{\max}-y^{\rm Fwad}_{\min}}\int_{y^{\rm
Fwad}_{\min}}^{y^{\rm Fwad}_{\max}} f(y,y_x,T_x)dy_x,
\end{align}
In Equation~\ref{eq.10}, $k$ ($1-k$) is the contribution fraction of the sources in
the backward (forward) rapidity region, and $[y^{\rm
Bwad}_{\min},y^{\rm Bwad}_{\max}]$ ($[y^{\rm Fwad}_{\min},y^{\rm
Fwad}_{\max}]$) is the rapidity shift range of the sources in the
backward (forward) rapidity region. For symmetric collisions, one
has $k=0.5$, $y^{\rm Bwad}_{\min}=-y^{\rm Fwad}_{\max}$, and
$y^{\rm Bwad}_{\max}=-y^{\rm Fwad}_{\min}$. Generally, both
$y^{\rm Bwad}_{\max}$ and $y^{\rm Fwad}_{\min}$ are approximately
equal to 0.

However, for experimental $y$ distribution of net-baryons, the
weight is obviously large in the very backward and forward
rapidity regions. Meanwhile, the weight decreases when the source
shifts to the central rapidity region. One may consider a
logarithmic Gaussian distribution obeyed by the weight.
Considering the rapidity shifts in the backward and forward
regions and a reflection in the two regions, one has the mutually
reflective two-logarithmic Gaussian distribution:
\begin{align}
\label{eq.11} f(y)=\frac{1}{N}\frac{dN}{dy}=&
\frac{k}{\sqrt{2\pi}\sigma_{\rm Bwad}} \int_{y^{\rm
Bwad}_{\min}}^{y^{\rm Bwad}_{\max}} \frac{1}{y_x-y^{\rm
Bwad}_{\min}}\exp\bigg\{-\frac{\ln\big[(y_x -y^{\rm
Bwad}_{\min})-y_{\rm Pk}^{\rm Bwad}\big]^2}{2\sigma_{\rm Bwad}^2}
\bigg\} f(y,y_x,T_x)dy_x \nonumber\\
&+ \frac{1-k}{\sqrt{2\pi}\sigma_{\rm Fwad}}\int_{y^{\rm
Fwad}_{\min}}^{y^{\rm Fwad}_{\max}} \frac{1}{y^{\rm
Fwad}_{\max}-y_x}\exp\bigg\{-\frac{\ln\big[(y^{\rm
Fwad}_{\max}-y_x)-y_{\rm Pk}^{\rm Fwad}\big]^2}{2\sigma_{\rm
Fwad}^2} \bigg\} f(y,y_x,T_x)dy_x,
\end{align}
where $\sigma_{\rm Bwad}$ ($\sigma_{\rm Fwad}$) and $y_{\rm
Pk}^{\rm Bwad}$ ($y_{\rm Pk}^{\rm Fwad}$) are the distribution
width and peak position in the backward (forward) rapidity region.
For symmetric collisions, one has $\sigma_{\rm Bwad}=\sigma_{\rm
Fwad}$ and $y_{\rm Pk}^{\rm Bwad}=-y_{\rm Pk}^{\rm Fwad}$.

In Equation~\ref{eq.11}, considering that the rapidity of the emission source can take any value within the dynamically allowed range, this seems to imply that the number of emission sources is infinite. However, in reality, our experience shows that fewer than 10 emission sources are needed to describe experimental data. Additionally, it should be noted that both Equation~(\ref{eq.3}) and Equation~(\ref{eq.11}) do not differentiate between the contributions of soft and hard processes in the rapidity distribution provided by the model. To distinguish the contributions of the two processes, two sets of parameters are required, each corresponding to one process, and their contributions are combined into the final rapidity distribution according to a certain proportion. The selection of more detailed rapidity parameters for soft and hard processes is beyond the scope of this article. This article only uses one set of parameters, which can be seen as a weighted average of two sets of parameters.

The chemical potential $\mu$ in $dN/dy$ and $f(y,y_x,T_x)$ can be
obtained by some methods. For example, one may use three
independent chemical potentials [baryon ($\mu_B$), electric charge
or isospin ($\mu_I$), and strangeness
($\mu_S$)]~\cite{39a,39b,39c} and related conserved charges to
obtain $\mu$~\cite{39d,39e,39f,39g}. Alternatively, one may use
the yield ratio~\cite{39h,39i} of negatively to positively charged
hadrons and source temperature to obtain $\mu$~\cite{39j,39k,39l}.
Different types of particles correspond to different $\mu$. This
work focuses on the distribution of net-baryons for which one has
empirically~\cite{39g}
\begin{equation}
\label{eq.12} \mu_B=\frac{1.3075}{1+0.288(\sqrt{s_{NN}}/\rm GeV)}
\,\, \rm (GeV)
\end{equation}
for baryons and $-\mu_B$ for anti-baryons. The constants (1.3075 and 0.288) in Equation~\ref{eq.12} may vary slightly in previous literature due to the fits for different statistical counts.

\begin{figure*}[htb!]
\begin{center}
\includegraphics[width=8.0cm]{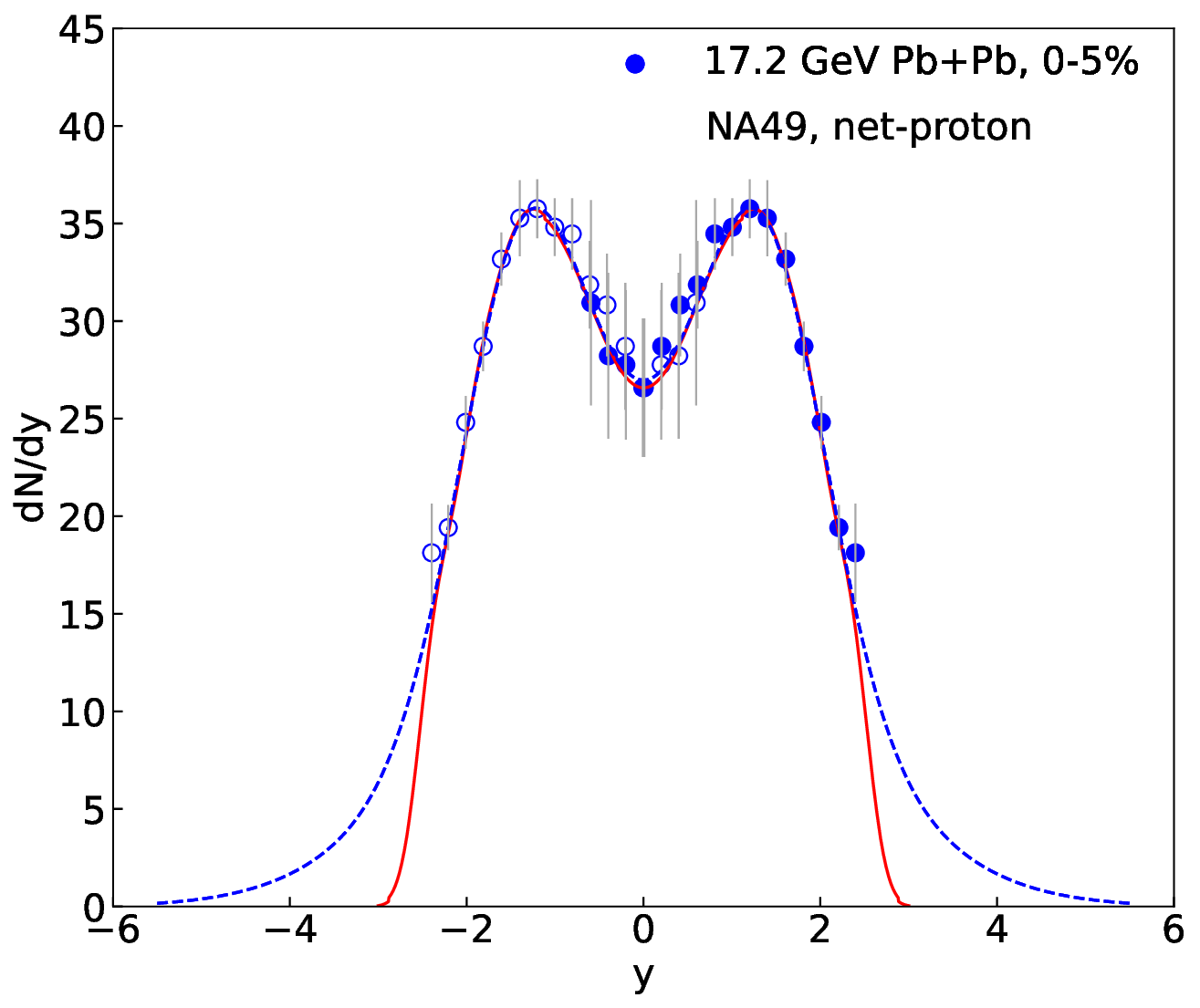}
\end{center}
\justifying\noindent {Figure 1. Rapidity density, $dN/dy$, of
net-protons produced in 0--5\% Pb-Pb collisions at 17.2 GeV. The
closed symbols represent experimental data measured by the NA49
Collaboration~\cite{39m}, and the open symbols are reflections of
the closed ones. The solid curves are our results fitted by the
multi-component distribution [Eq.~(\ref{eq.11})], while the dashed
curves are our results fitted by the three-Gaussian distribution
[Eq.~(\ref{eq.5})].}
\end{figure*}

\begin{figure*}[htb!]
\begin{center}
\includegraphics[width=8.0cm]{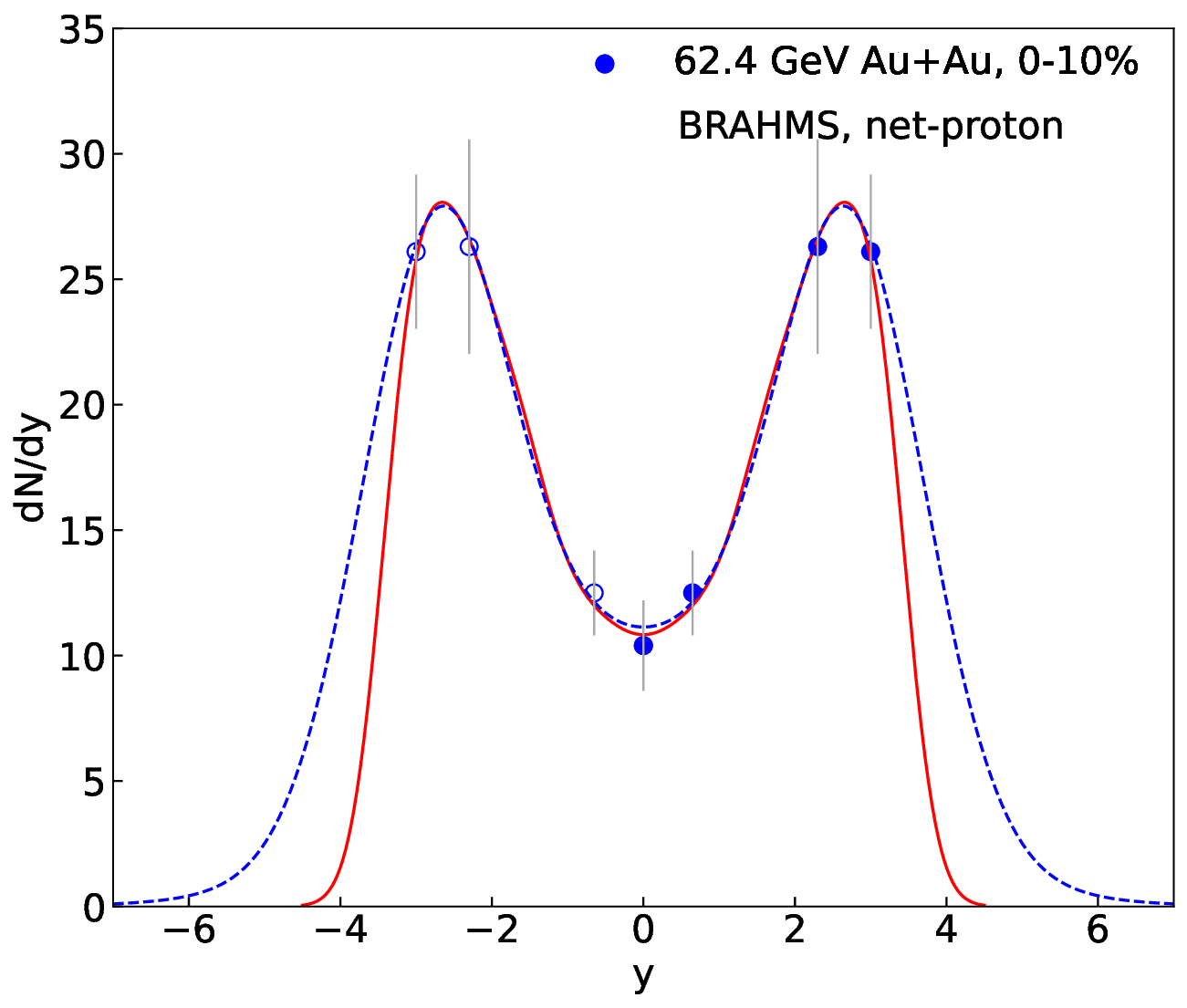}
\end{center}
\justifying\noindent {Figure 2. Rapidity density of net-protons
produced in 0--10\% Au-Au collisions at 62.4 GeV. The closed and
open symbols represent experimental data measured by the BRAHMS
Collaboration~\cite{39n,39o} and corresponding reflections
respectively. The solid and dashed curves are our results fitted
by the multi-component distribution [Eq.~(\ref{eq.11})] and the
three-Gaussian distribution [Eq.~(\ref{eq.5})] respectively.}
\end{figure*}

\begin{figure*}[htb!]
\begin{center}
\includegraphics[width=8.0cm]{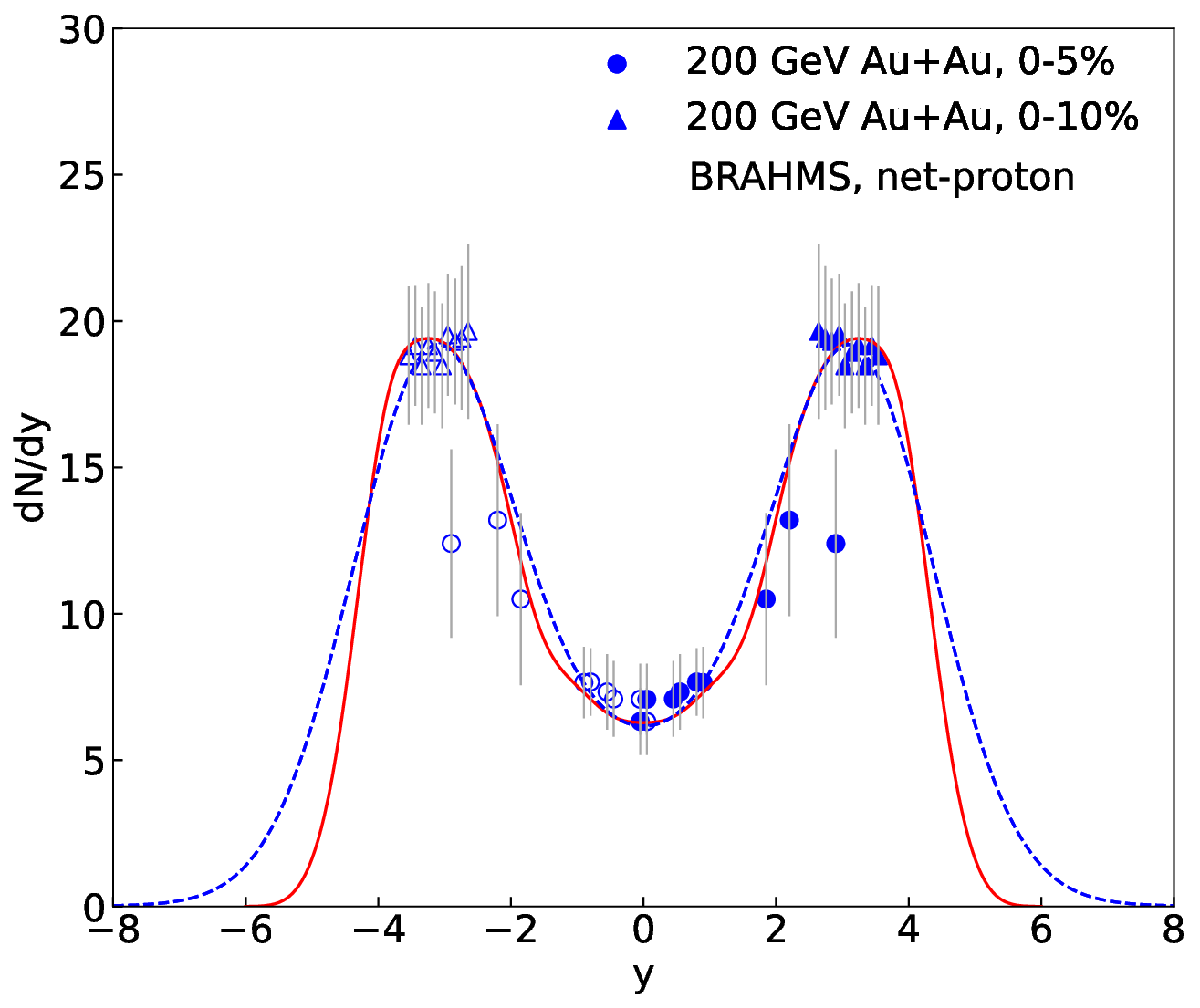}
\end{center}
\justifying\noindent {Figure 3. Rapidity density of net-protons
produced in 0--5\% (0--10\%) Au-Au collisions at 200 GeV. The
closed and open symbols represent experimental data measured by
the BRAHMS Collaboration~\cite{39o,39p} and corresponding
reflections respectively. The solid and dashed curves are our
results fitted by the multi-component distribution
[Eq.~(\ref{eq.11})] and the three-Gaussian distribution
[Eq.~(\ref{eq.5})] respectively.}
\end{figure*}

\section{Rapidity distribution of net-protons}

Based on previous investigation, one knows that $T_x$ in central
rapidity region is larger than that in backward and forward
rapidity regions. For collisions at GeV, we may assume generally
that $T_x\approx0.2\sim0.4$ GeV in central rapidity region (with
$y_x=y^{\rm Bwad}_{\max}$ or $y^{\rm Fwad}_{\min}$) and
$T_x\approx0.05\sim0.15$ GeV in backward and forward rapidity
regions (with $y_x=y^{\rm Bwad}_{\min}$ and $y^{\rm Fwad}_{\max}$
respectively). From central to backward and forward rapidity
regions, $T_x$ is assumed to decrease linearly.

Figures 1--3 present rapidity density ($dN/dy$) of net-protons
produced in 0--5\% Pb-Pb collisions at $\sqrt{s_{NN}}=17.2$ GeV,
0--10\% Au-Au collisions at $\sqrt{s_{NN}}=62.4$ GeV, and 0--5\%
(0--10\%) Au-Au collisions at $\sqrt{s_{NN}}=200$ GeV,
respectively. The closed symbols represent experimental data
measured by the NA49~\cite{39m} and BRAHMS
Collaborations~\cite{39n,39o,39p}, and the open symbols are
reflections of the closed ones. The solid curves are our results
fitted by the multi-component distribution [Eq.~(\ref{eq.11})],
while the dashed curves are our results fitted by the
three-Gaussian distribution [Eq.~(\ref{eq.5})].

It should be noted that in practical fitting, in order to avoid
the excessive use of computers for normalizing too many
$f(y,y_x,T_x)$, we only selected a limited number of $y_x$ and
performed logarithmic Gaussian distribution fitting on the limited
$y_x$ values and their weights, obtaining the corresponding widths
and peak positions.

From Figures 1--3 one can see that both distributions
[Eqs.~(\ref{eq.11}) and (\ref{eq.5})] approximately describe the
trend of rapidity density of net-protons in available data region
in central $AA$ collisions at high energy. Comparing with the
multi-component distribution [Eq.~(\ref{eq.11})], the
three-Gaussian distribution [Eq.~(\ref{eq.5})] overestimates the
yield of net-protons in the backward and forward rapidity regions.

The dependence trends of related free parameters on center-of-mass
energy are analyzed by us. Figure 4 displays the dependence of
$T_x$ versus $\sqrt{s_{NN}}$ for the sources at the central
rapidity which has the maximum $T_x$ ($T_{x}|_{\max}$), and for
the sources at the minimum rapidity which has the minimum $T_x$
($T_{x}|_{\min}$). In Figure 4, the squares and circles represent
$T_{x}|_{\max}$ and $T_{x}|_{\min}$, respectively. Figure 5 gives
the dependence of (a) $\sigma_{\rm Bwad}$ versus $\sqrt{s_{NN}}$
(circles) and (b) $|y_{\rm Pk}^{\rm Bwad}|$ (squares),
$|y_{\min}^{\rm Bwad}|$ (triangles), and $|y_{\max}^{\rm Bwad}|$
(asterisks) versus $\sqrt{s_{NN}}$ obtained from the fit of
mutually reflective two-logarithmic Gaussian distribution
[Eq.~(\ref{eq.11})]. Figure 6 presents the dependence of (a)
$\sigma_{C}$ and $\sigma_T$ versus $\sqrt{s_{NN}}$ (open and
closed circles) and (b) $|y_C|$ and $|y_T|$ versus $\sqrt{s_{NN}}$
(open and closed squares) obtained from the fit of three-Gaussian
distribution [Eq.~(\ref{eq.5})].

One can see from Figures 4--6 that most parameters increase
approximately linearly with the increase of logarithmic
center-of-mass energy, though $y_{\max}^{\rm Bwad}$ and $y_C$ are
always equal to 0. Particularly, $T_x$ in the central rapidity
region is larger than that in the backward rapidity region. This
also implies that $\langle p_T\rangle$ in the central rapidity
region is larger than that in the backward rapidity region.

\begin{figure*}[htb!]
\begin{center}
\includegraphics[width=8.0cm]{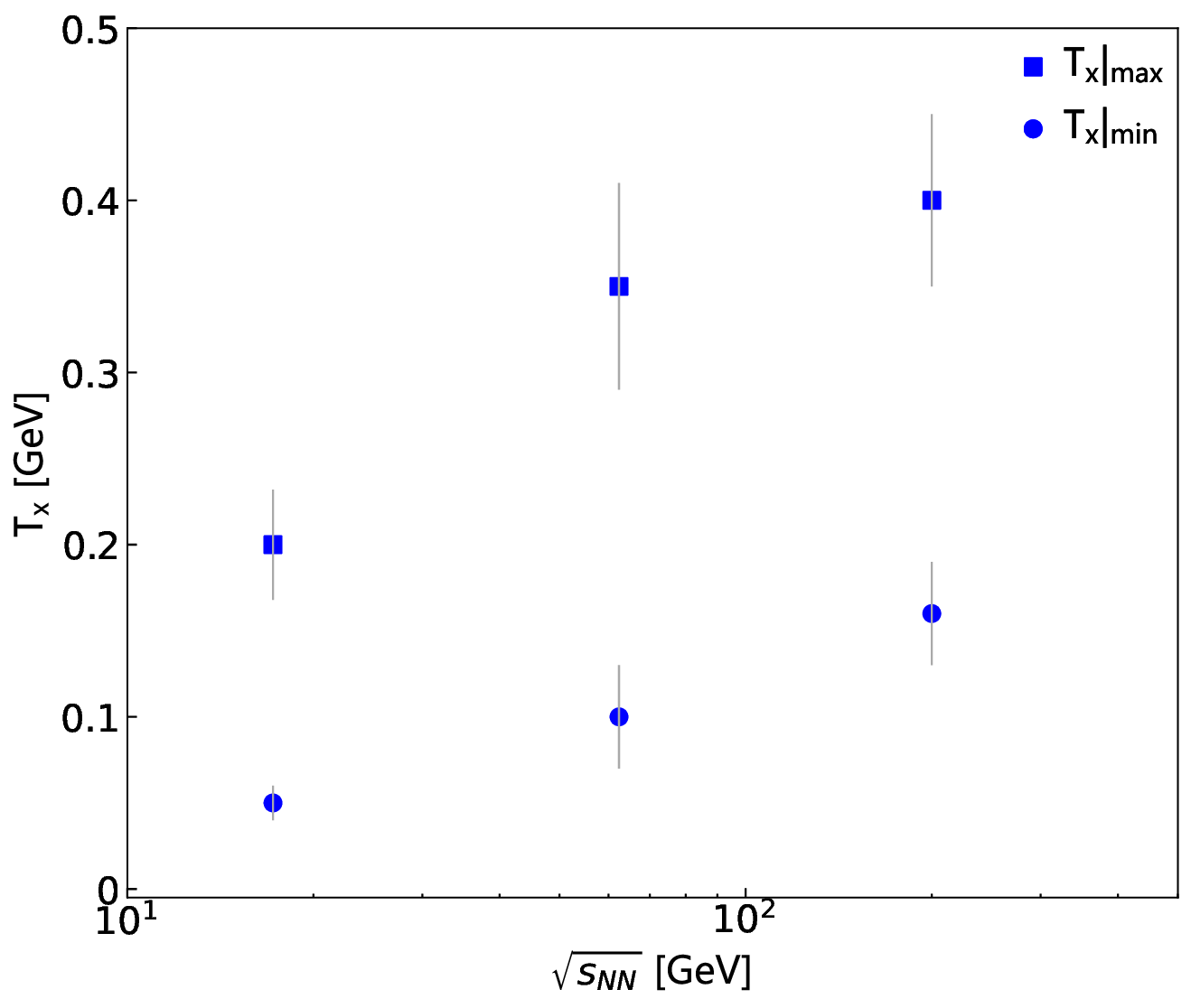}
\end{center}
\justifying\noindent {Figure 4. Dependence of $T_x$ versus
$\sqrt{s_{NN}}$. For the sources at the central rapidity, $T_x$ is
taken to be the maximum $T_{x}|_{\max}$ and is represented by the
squares. For the sources at the minimum rapidity, $T_x$ is taken
to be the minimum $T_{x}|_{\min}$ and is represented by the
circles. These $T_x$ are used in Eq.~(\ref{eq.11}).}
\end{figure*}

\begin{figure*}[htb!]
\begin{center}
\includegraphics[width=15.0cm]{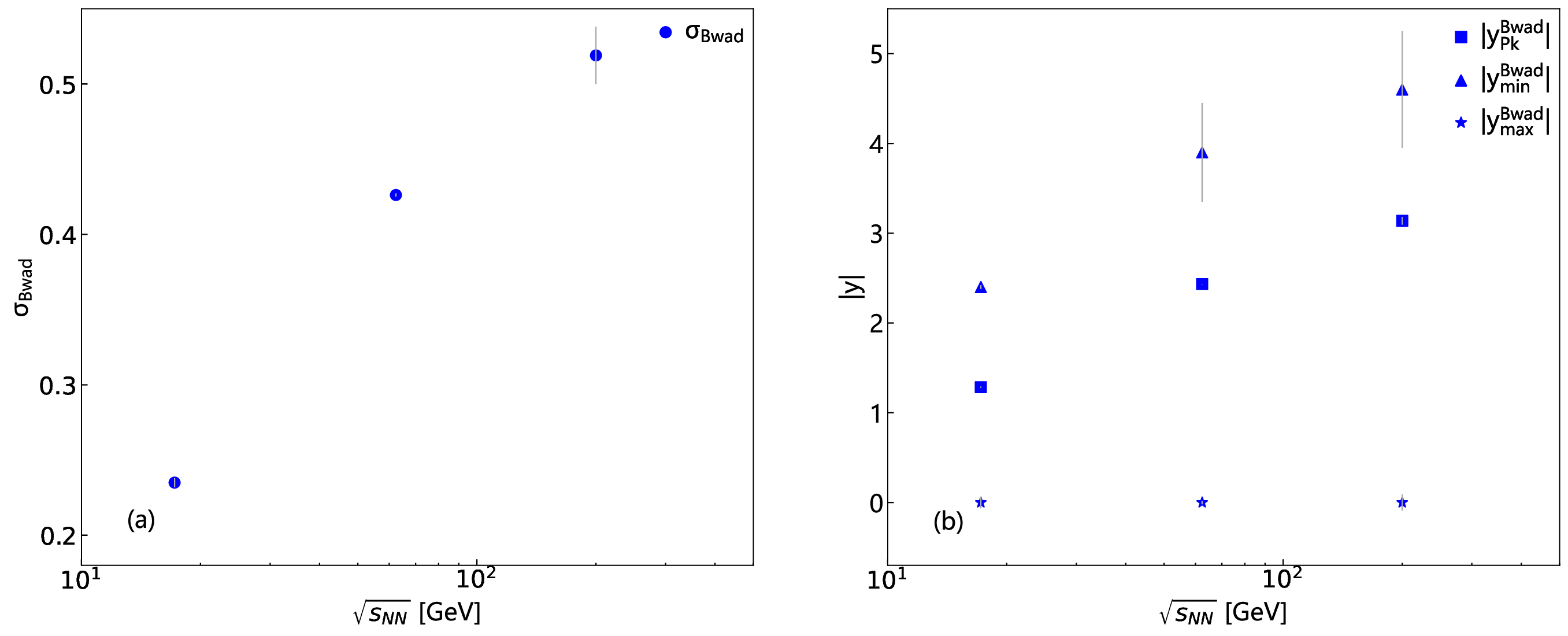}
\end{center}
\justifying\noindent {Figure 5. (a) Dependence of $\sigma_{\rm
Bwad}$ versus $\sqrt{s_{NN}}$ (circles). (b) Dependence of
$|y_{\rm Pk}^{\rm Bwad}|$ (squares), $|y_{\min}^{\rm Bwad}|$
(triangles), and $|y_{\max}^{\rm Bwad}|$ (asterisks) versus
$\sqrt{s_{NN}}$. All parameter values are obtained from the fit by
Eq.~(\ref{eq.11}).}
\end{figure*}

\begin{figure*}[htb!]
\begin{center}
\includegraphics[width=15.0cm]{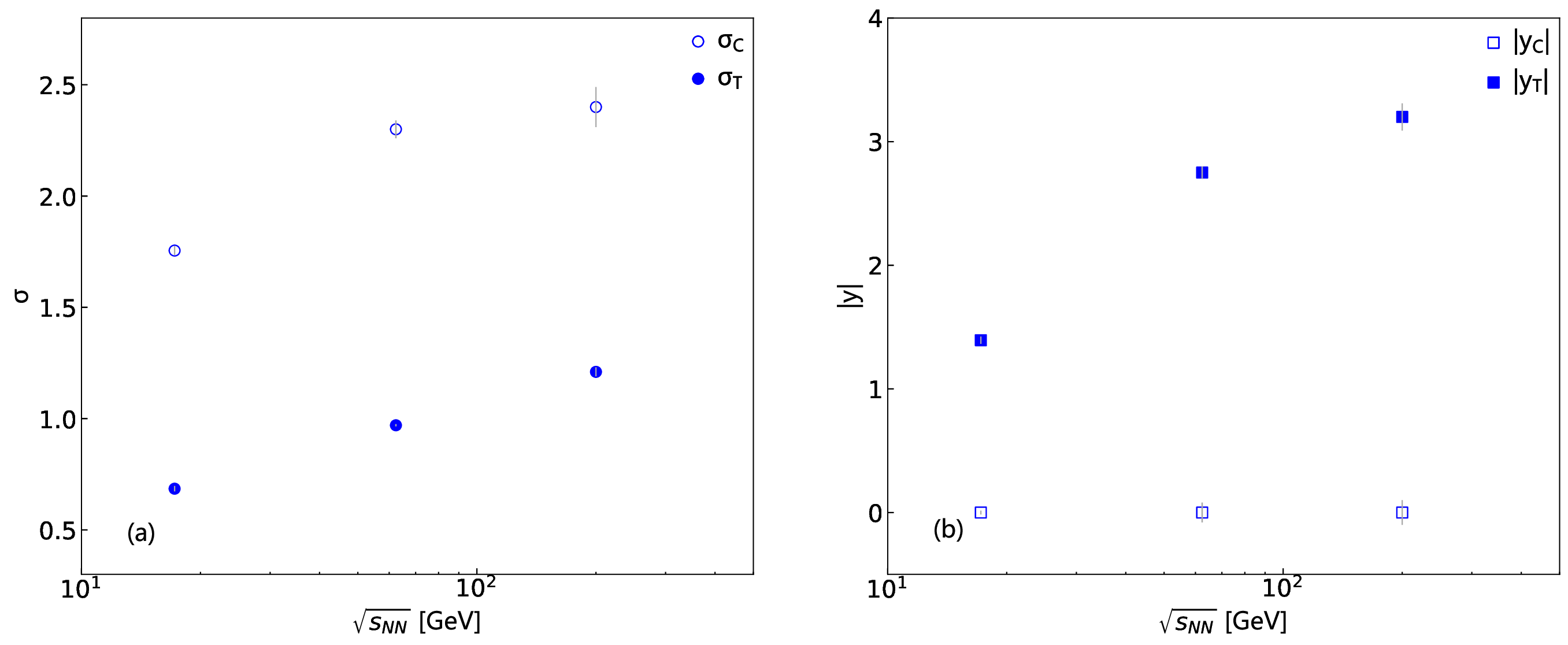}
\end{center}
\justifying\noindent {Figure 6. (a) Dependence of $\sigma_{C}$
(open circles) and $\sigma_T$ (closed circles) versus
$\sqrt{s_{NN}}$. (b) Dependence of $|y_C|$ (open squares) and
$|y_T|$ (closed squares) versus $\sqrt{s_{NN}}$. All parameter
values are obtained from the fit by Eq.~(\ref{eq.5}).}
\end{figure*}

\begin{figure*}[htb!]
\begin{center}
\includegraphics[width=8.0cm]{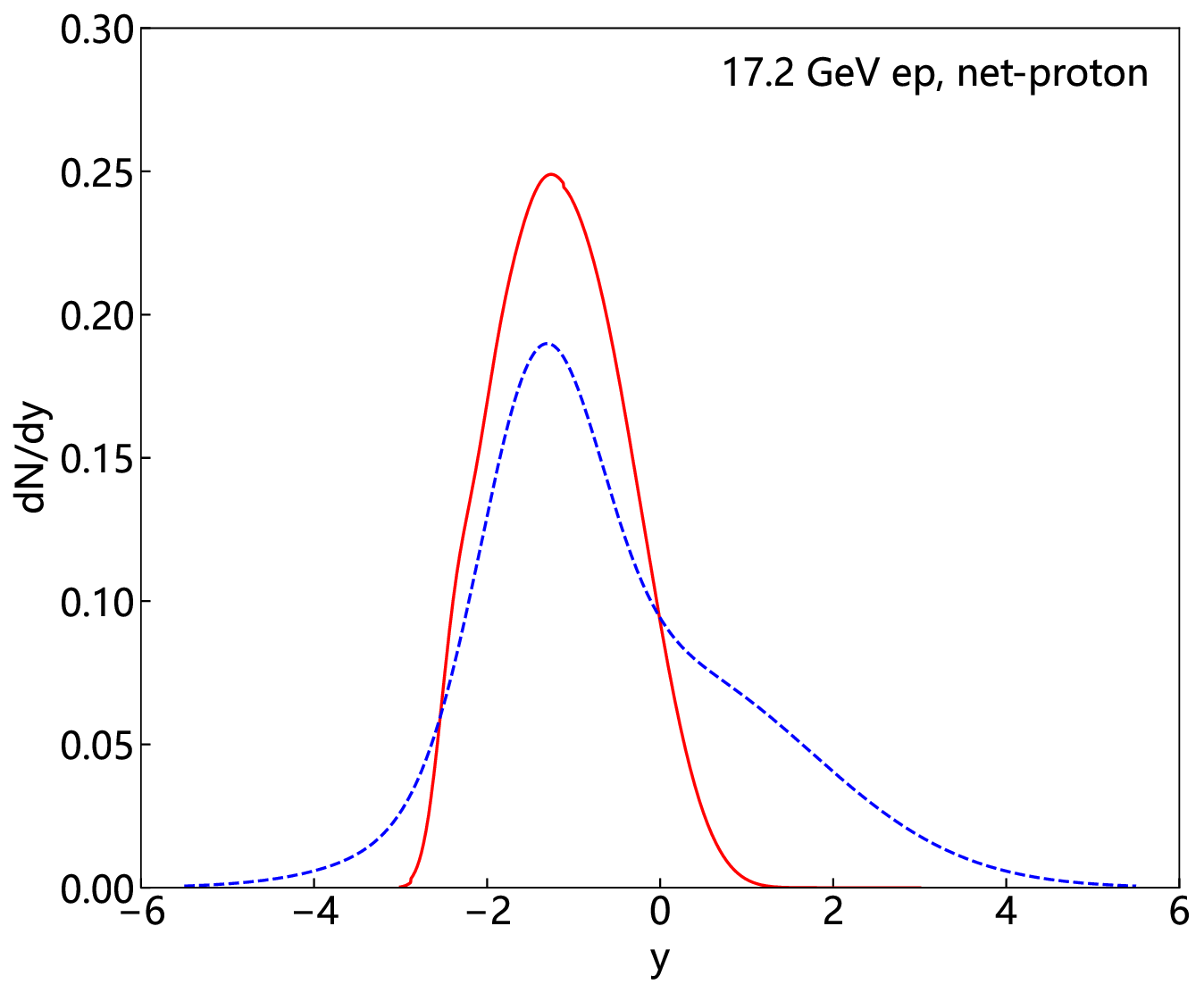}
\end{center}
\justifying\noindent {Figure 7. Predicted rapidity density of
net-protons produced in $ep$ collisions at 17.2 GeV. The solid and
dashed curves represent the results from the multi-component
distribution [Eq.~(\ref{eq.11})] and the three-Gaussian
distribution [Eq.~(\ref{eq.5})] respectively, where the direct
contribution of projectile $e$ does not exist.}
\end{figure*}

\begin{figure*}[htb!]
\begin{center}
\includegraphics[width=8.0cm]{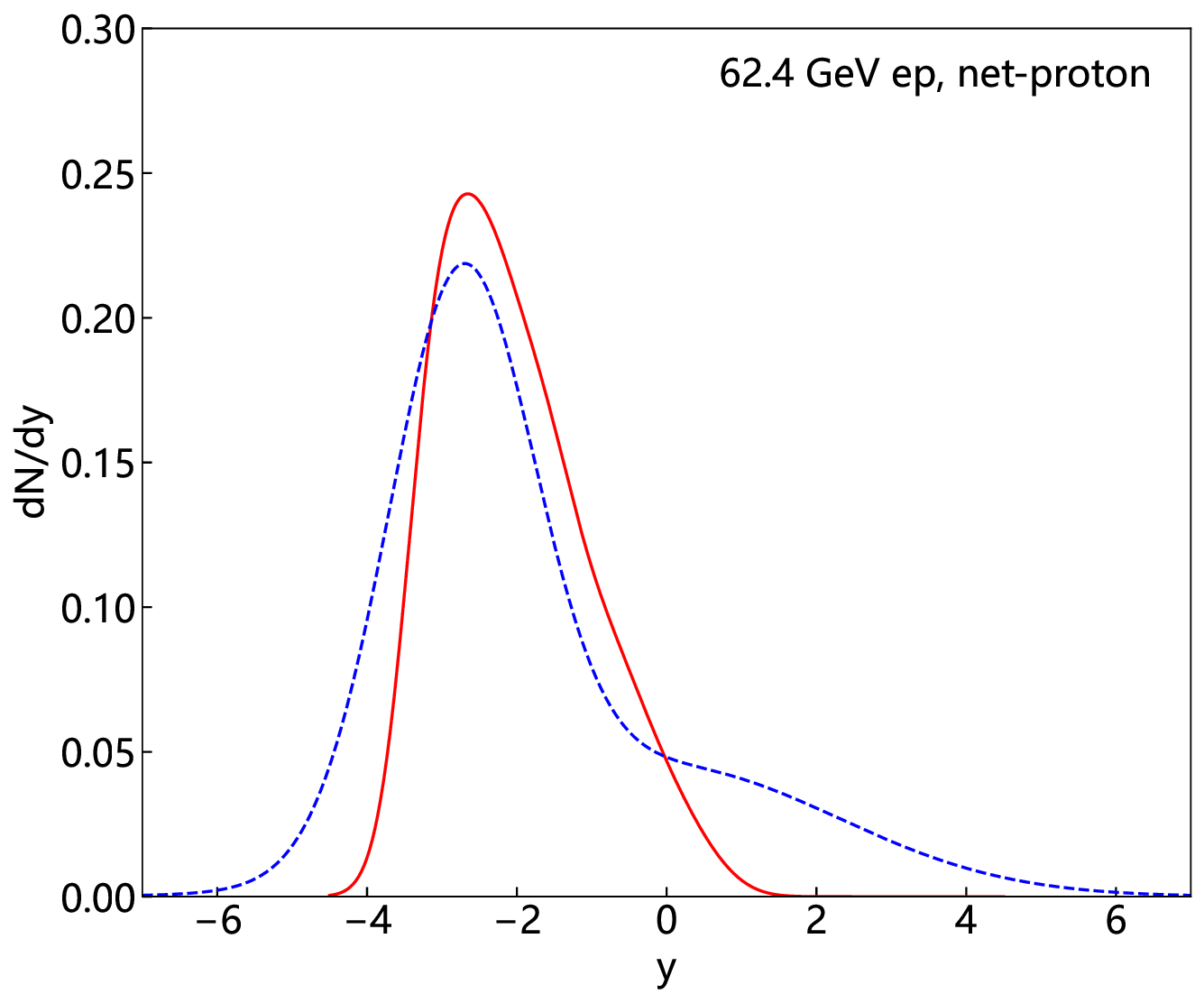}
\end{center}
\justifying\noindent {Figure 8. Predicted rapidity density of
net-protons produced in $ep$ collisions at 62.4 GeV. The solid and
dashed curves represent the results from the multi-component
distribution [Eq.~(\ref{eq.11})] and the three-Gaussian
distribution [Eq.~(\ref{eq.5})] respectively, where the direct
contribution of projectile $e$ does not exist.}
\end{figure*}

\begin{figure*}[htb!]
\begin{center}
\includegraphics[width=8.0cm]{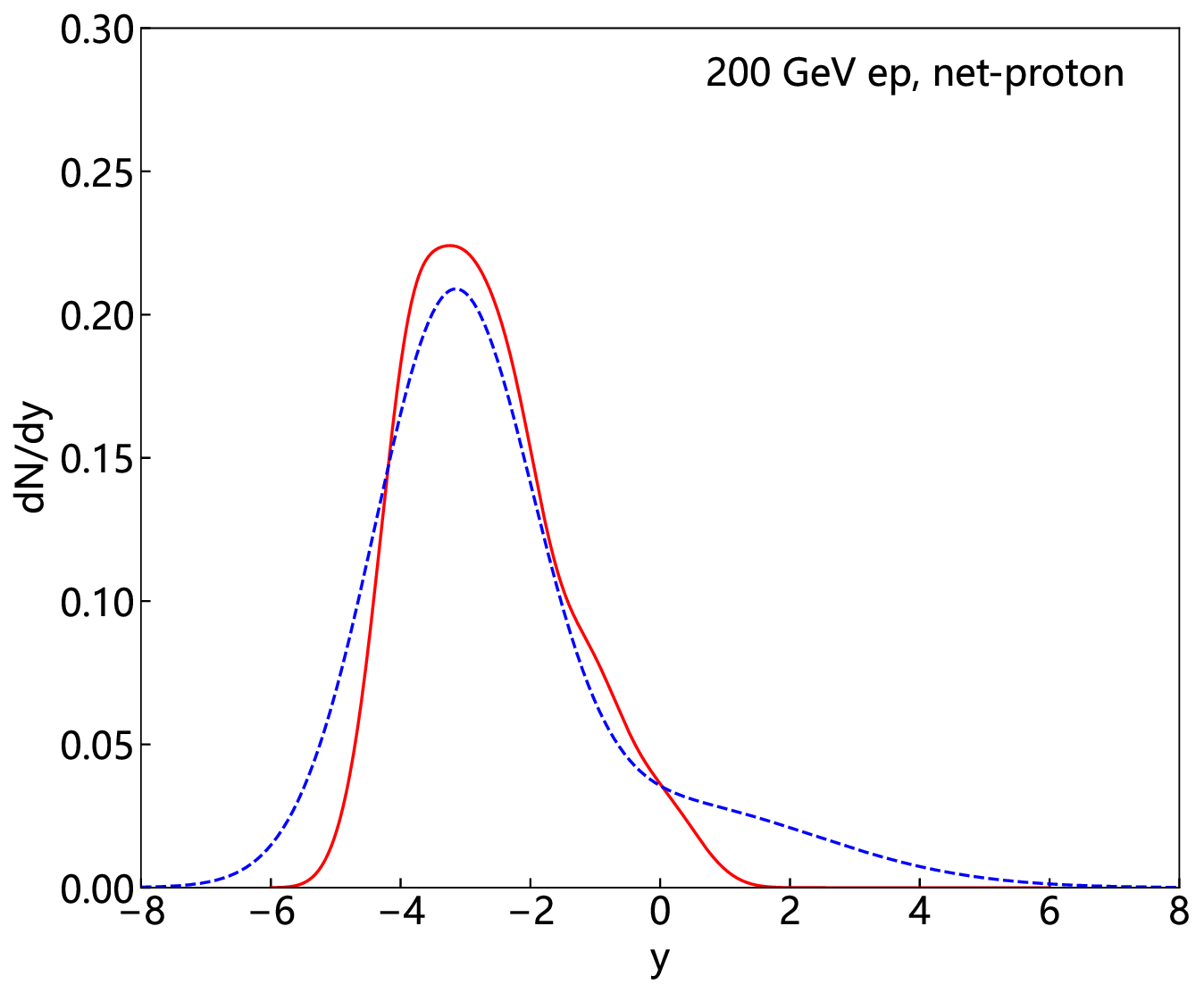}
\end{center}
\justifying\noindent {Figure 9. Predicted rapidity density of
net-protons produced in $ep$ collisions at 200 GeV. The solid and
dashed curves represent the results from the multi-component
distribution [Eq.~(\ref{eq.11})] and the three-Gaussian
distribution [Eq.~(\ref{eq.5})] respectively, where the direct
contribution of projectile $e$ does not exist.}
\end{figure*}

After giving up the residual nucleus in $eA$ collisions and not
considering it, the rapidity density of net-protons produced in
$eA$ collisions is similar to that in $ep$ collisions. To
understand the trend of rapidity distribution of net-protons
produced in $ep$ collisions at the EIC, we may predict some
results at different collision energies due to the trends of free
parameters.

As examples, Figures 7--9 show the predicted rapidity distribution
of net-protons from $ep$ collisions at center-of-mass energy
$\sqrt{s}=7.2$, 62.4, and 200 GeV, respectively. The projectile
$e$-going (target $p$-going) direction points towards the forward
(backward) rapidity region. The solid and dashed curves represent
the results from the multi-component distribution
[Eq.~(\ref{eq.11})] and the three-Gaussian distribution
[Eq.~(\ref{eq.5})] respectively. We would like to point out that
the direct contribution of projectile $e$ to the yield of
net-protons does not exist. All results are from the direct
contribution of target $p$. Even for the curves in the forward
rapidity region, they also represent the direct contributions of
target $p$, for which the sources are in the backward and central
rapidity regions.

One can see from Figures 7--9 that, comparing with the
multi-component distribution [Eq.~(\ref{eq.11})], the
three-Gaussian distribution [Eq.~(\ref{eq.5})] over predicts the
yields of net-protons in the backward and forward rapidity
regions. The constraint of beam rapidity indicates that the
multi-component distribution [Eq.~(\ref{eq.11})] is more accurate
than the three-Gaussian distribution [Eq.~(\ref{eq.5})]. The above
fit and prediction are based on the relative large $T_x$ in
central rapidity region. This is exactly what the valence
quark-stopping scenario requires.

\section{Discussion on what carries the baryon number}

In soft excitation processes induced by principal participating sea quarks and gluons, gluon junctions composed of low-momentum gluons interacting primarily with the soft parton field typically localize in the central rapidity region~\cite{17,18} when there is no penetrability between the projectile and target. However, in high-energy collisions characterized by strong penetrability, gluon junctions (and the baryons they subsequently form) can deviate from the central region, stopping in either the backward or forward rapidity regions. This deviation is most pronounced in the extreme limit of soft processes. In hard scattering processes induced by pairs of principal participating partons, valence quarks carry a significant fraction of the baryon's momentum and undergo short interaction times. In the absence of stopping effects, these quarks would be expected to end up in the backward or forward rapidity regions~\cite{17,18}. However, due to the substantial stopping power inherent in high-energy collisions involving valence quarks, they instead terminate their trajectories in the central rapidity region with high probability, even in the extreme limit of hard processes.

As previously discussed, during collisions between a projectile electron \(e\) and a target nucleon \(N\) within a nucleus \(A\), net-baryons are distributed across both the backward and central rapidity regions. Notably, net-baryons do not appear in the forward rapidity region, where no leading nucleons originate from the projectile itself. In addition to baryons, other particles can also be produced as a result of \(eA\) collisions. Following particle production events, any remaining nucleons may form an excited nucleus that can fragment into various nuclear fragments. In this work, we will not delve further into discussions regarding other particles or nuclear fragments, as they do not influence our assessment of what carries the baryon number.

The first component in Eq.~(\ref{eq.3}) describes \(eN\) scattering involving sea quarks, gluons, and a lepton. This relatively non-violent soft excitation interaction generates a low-temperature baryon source~\cite{29}, with net-baryon distributions determined by two distinct stopping scenarios. In the valence quark-stopping scenario, at high energies, non-principal participating valence quarks exhibit strong penetrability, leading net-baryons to concentrate in the backward rapidity region (where the low-temperature source is localized). In the gluon junction-stopping scenario, the gluon junction loses energy due to strong stopping power, resulting in net-baryons being primarily distributed in the central rapidity region (where the low-temperature source is localized). Notably, the spatial separation of net-baryon distributions between the backward and central rapidity regions allows for clear differentiation between the two scenarios via measurable source temperature differences.

The second component in Eq.~(\ref{eq.3}) describes \(eN\) scattering involving a principal participating parton and a lepton. This corresponds to a high-temperature baryon source generated by violent hard-scattering collisions~\cite{29}. In the valence quark-stopping scenario, within our model framework, net-baryons are expected to concentrate in the central rapidity region if the hard process is induced by a principal participating valence quark and a lepton. This prediction relies on the assumption that valence quarks undergo sufficient multiple scatterings during the hard process to lose a significant fraction of their rapidity. Alternatively, net-baryons shift to the backward rapidity region if the process involves a principal participating gluon and a lepton; in this case, the valence quarks, as non-principal participants, participate throughout the entire process and are left behind in the backward rapidity region. In the gluon junction-stopping scenario, net-baryons localize in the backward rapidity region if the hard process is induced by a principal participating valence quark and a lepton, or they concentrate in the central rapidity region if the process involves a principal participating gluon and a lepton. Notably, both scenarios predict net-baryons in the backward and central rapidity regions in hard scattering processes, making them indistinguishable based solely on the overall rapidity range of net-baryon distributions. More refined observables, such as the yield ratio between central and backward regions or net-baryon-charge correlations, might offer additional discriminating power.

Based on the above discussion, the multi-source thermal model effectively describes both soft excitation and hard scattering processes~\cite{28,29}, clarifying the two-component distribution characteristics of particle multiplicities, transverse momenta, and rapidities. Key observations indicate that soft excitation produces particles spanning backward to forward rapidity regions, associated with a low-temperature emission source. Hard scattering generates particles predominantly concentrated in the central rapidity region, with slight extensions into the backward and forward rapidity regions, corresponding to a high-temperature source. Central rapidity particles originate from both low- and high-temperature sources, while non-central particles are dominated by the low-temperature source, with minimal high-temperature contribution.

This framework links particle production mechanisms to their rapidity distributions, providing a basis for testing baryon number carriers via EIC \(eA\) collision experiments. If valence quarks carry the baryon number, soft excitation produces net-baryons in the backward rapidity region (low-temperature source), while hard scattering concentrates them in the backward and central regions (high-temperature source). This predicts higher temperatures or larger \(\langle p_T\rangle\) in the central rapidity region. If gluon junctions carry the baryon number, soft excitation localizes net-baryons in the central rapidity region (low-temperature source), while hard scattering shifts them to the backward and central regions (high-temperature source). This predicts lower temperatures or smaller \(\langle p_T\rangle\) in the central rapidity region.

Experimentally, baryon \(p_T\) distributions in the backward and central rapidity regions can be measured and fitted with a two-component Erlang distribution to derive \(\langle p_T \rangle\) (a proxy for temperature). Alternatively, direct estimates of \(\langle p_T \rangle\) from raw data can be used. It should be noted that \(\langle p_T \rangle\) is also influenced by radial flow. This qualitative criterion implicitly assumes that the radial flow contribution does not differ significantly between the two rapidity regions; a rigorous quantitative application would require decomposing the thermal and flow contributions using hydrodynamic or transport models. A smaller \(\langle p_T \rangle\) in the backward region compared to the central region supports the valence quark-stopping scenario, while a larger \(\langle p_T \rangle\) in the backward region favors the gluon junction-stopping scenario.

After excluding leading nucleon contributions, baryons and mesons produced in collision systems should exhibit similar rapidity dependence, as they originate from the same emission source. Even in a small system like \(eN\) scattering, we assume that instantaneous local equilibrium is achieved and that the baryons and mesons come from the same emission source. This implies analogous rapidity-dependent behavior between baryons and common charged particles. Our previous work~\cite{40,41,42}, the current study, and related research~\cite{43} consistently show that the emission source temperature or charged particle \(\langle p_T\rangle\) generally decreases with increasing \(|y|\). This observation supports the scenario where valence quarks act as baryon number carriers: net-baryons from low-intensity soft processes dominate the backward rapidity region, while those from high-intensity hard processes concentrate in the central rapidity region. Of course, these works mainly focus on the GeV energy region and fail to make predictions for the TeV energy region, where the situation will be different, or even opposite. Based on comprehensive analysis, we preliminarily conclude that valence quarks are indeed baryon number carriers at GeV energies, which is a natural prediction of our multi-source thermal model~\cite{28,29}, which will be further tested in future EIC experiments operating at GeV energies.

For upcoming EIC \(eA\) collision experiments, determining the correct stopping scenario requires only measuring and comparing charged particle \(\langle p_T\rangle\) in the backward and central rapidity regions. If the central-region \(\langle p_T\rangle\) exceeds that in the backward region, the valence quark-stopping scenario is favored (hard-process baryons concentrate in central rapidity). If the central-region \(\langle p_T\rangle\) is smaller, the gluon junction-stopping scenario is supported (hard-process baryons dominate backward rapidity). If there is no significant difference between the two \(\langle p_T\rangle\) values within the uncertainty range, a definitive judgment cannot be made.

\section{Discussion on the consistency of the physical picture}

In the above discussion, we examined two Wilson line (or gluon flux-tube) topologies: the \(\triangle\)-shaped and Y-shaped configurations. These structures represent two extreme scenarios: the \(\triangle\)-shaped topology corresponds to a highly excited state of the baryon, while the Y-shaped topology corresponds to its static ground state. In the \(\triangle\)-shaped topology, the three Wilson lines vibrate without interconnections, whereas in the Y-shaped topology, they are fully joined at a single central junction, effectively forming three sets of pairwise links. Driven by inherent vibrations, connection formation, and link breakage, the two topologies can interconvert. The most probable state is thus a dynamic mixture of partial connections and fractures, rather than a strict adherence to either extreme configuration.

This work aims to explore possible mechanisms of baryon number transfer in high-energy collisions and proposes a scenario dominated by the \(\triangle\)-shaped topology as an alternative to the classical Y-shaped interpretation~\cite{15,16,16aa}. We clarify that this does not negate the effectiveness of the Y-shaped topology in describing the baryon's static ground state. Indeed, lattice QCD studies provide solid evidence for Y-shaped flux-tube configurations within baryons at the hadronic scale. For instance, Refs.~\cite{15,16,16aa} demonstrate through precise calculations of hundreds of distinct configurations that the three-quark potential is well described by a Coulomb term plus a Y-shaped linear potential, and directly observe the formation of Y-shaped flux-tubes.

In the present work, we clarify that our results are not necessarily contradictory to important lattice QCD findings, as the two studies focus on distinct subjects. Classical lattice QCD calculations primarily investigate the potential of heavy subsystems in static ground states, where the Y-shaped topology represents the lowest-energy flux-tube configuration~\cite{15,16,16aa}. In contrast, our work focuses on the dynamic mechanisms of high-energy scattering processes, particularly baryon number transfer in unexplored kinematic regimes which is a process that may involve highly non-stationary states, excited states, or specific kinematic conditions. We emphasize that the two topologies are complementary, each describing distinct aspects of baryon structure and dynamics under different physical conditions.

Moreover, the three-quark system can exist in gluon-excited states with excitation energies up to about 1 GeV. This suggests that during dynamic processes, the effective configuration of the system may deviate from the ground-state Y-shaped topology. The \(\triangle\)-shaped topology could correspond to a specific excited-state configuration or a dominant transfer pathway in certain dynamic processes, without invalidating the Y-shaped ground-state topology. The two topologies likely describe different aspects of the same physical system under varying energies, time scales, or observational conditions. Lattice QCD evidence addresses the question of what the lowest-energy shape of a static baryon is, while our model-based work attempts to answer how baryon number flows under the specific conditions of high-energy collisions. These two lines of inquiry complement each other, contributing to a more complete physical picture.

We note that attributing the enhanced high-energy or high-``temperature'' components observed at mid-rapidity to ``valence quark-stopping'' appears to contradict the simple kinematic picture, which traditionally holds that high Bjorken-\(x\) valence quarks remain closer to the beam direction due to strong Lorentz contraction. In fact, the ``valence quark-stopping'' we refer to describes an extreme hard-scattering scenario. The probability of a valence quark losing a significant fraction of its rapidity is close to zero. The traditional picture of high Bjorken-\(x\) valence quarks moving forward corresponds precisely to the soft excitation processes discussed in our work. In high-energy collisions, the proportion of soft excitation processes is significantly higher than that of hard scattering processes.

In our previous work~\cite{73}, we explored various possible energy loss fractions of leading protons in proton-induced nuclear collisions at 200 GeV/\(c\) which is a beam momentum corresponding approximately to a momentum transfer scale of \(Q\sim5\) GeV/\(c\) in inelastic scattering. In free-space proton-nucleon collisions, the most experimentally realistic scenario involves the incident proton losing 50\% of its energy. This energy loss fraction is carried by gluons, i.e., \(x_{g}=0.5\), as the energy is consumed in particle production via soft excitation processes dominated by gluonic interactions. The remaining 50\% of the proton's energy is roughly split equally between sea quarks and valence quarks, i.e., \(x_{sq}=0.25\) and \(x_{vq}=0.25\), assuming comparable populations of the two quark types. These energy fractions align with QCD-based predictions at similar momentum transfer scales~\cite{73a,73b}. Following the collisions, the gluons, sea quarks, and valence quarks in the proton reestablish dynamic equilibrium.

Correlation analysis between charge stopping and baryon stopping represents a core frontier method for distinguishing baryon number carriers (valence quarks vs. gluon junctions). In the traditional QCD framework, both baryon number and charge are carried by valence quarks, so their stopping behaviors in the central rapidity region should be closely correlated. In contrast, if baryon number is primarily carried by a non-perturbative, electrically neutral Y-shaped gluon field configuration (i.e., a gluon junction), charge transport and baryon number transport may become partially decoupled. Therefore, studying the correlation between net-charge yield and net-baryon yield in the central rapidity region is crucial for probing the microscopic mechanisms of baryon transport.

The correlation between net-baryon production and net-charge production at mid-rapidity can be investigated in different collision systems~\cite{16a,74}. We define two stopping parameters: the baryon stopping parameter \(\alpha_B\) (the ratio of net-baryon production to the initial total baryon number) and the charge stopping parameter \(\alpha_Q\) (the ratio of net-charge production to the initial total charge). Research has not yet revealed a significant and definitive correlation between \(\alpha_Q\) and \(\alpha_B\) across various systems and collision energies~\cite{16a}. However, a positive correlation between \(\alpha_Q\) and \(\alpha_B\) is qualitatively consistent with benchmark predictions from the ultra-relativistic QMD (UrQMD) model~\cite{74a,74b,74c}, which adopts valence quark transport as the primary stopping mechanism. The model treats valence quark transport as the dominant pathway for charge and baryon stopping, and its simulations also exhibit the general trend of charge stopping exceeding baryon stopping.

More importantly, this strong correlation provides robust support for the scenario in which valence quarks act as the primary carriers of baryon number. If baryon number were primarily transported by electrically neutral gluon junctions, there would be no inherent, direct correlation between charge and baryon stopping, making the observed strong linear relationship difficult to explain. Our analysis shows that the formation and evolution of matter at mid-rapidity is highly consistent with a transport scenario dominated by valence quark degrees of freedom. The correlated variations in net-charge and net-baryon yields reveal a physical picture in which baryon number and charge are transported together in the early stages of collisions, which is a key link in understanding valence quark transport dynamics.

We also note that charge stopping is generally more pronounced than baryon stopping, a phenomenon that may indicate contributions to charge transport from processes beyond valence quark transport, such as soft interactions involving gluons or sea quarks. Additionally, factors like the asymmetry in rapidity distributions between strange and anti-strange quarks may influence the baryon-to-charge ratio. Future work should seek to clarify the relative contributions of different mechanisms at a more refined level, incorporating richer observational constraints such as isospin effects and strangeness conservation.

Indeed, incorporating charge stopping into the analytical framework significantly enhances our ability to distinguish between baryon number transport mechanisms (valence quarks vs. gluon junctions). Our analysis indicates that the strong observed correlation between charge stopping and baryon stopping tends to support the scenario in which valence quarks act as the common carrier of both baryon number and charge in high-energy collisions. This finding provides important experimental clues for understanding the dynamical role of valence quarks in the evolution of high-temperature, high-density systems.

Before summarizing, we clarify that the multi-component model used in this study differs in formalism from the core prediction of the gluon junction-stopping mechanism based on Regge theory, which states that the net-baryon cross section decreases exponentially with rapidity loss~\cite{74d}. In regions of high rapidity loss, Regge theory-based models yield smaller net-baryon cross sections, whereas our multi-component model, each component being grounded in Landau hydrodynamics~\cite{31a,31e}, fits this region by adjusting weight factors through a log-Gaussian distribution. By tuning these weight factors, we expect the multi-component model's fitting results to align with those of Regge theory-based models.

Furthermore, the terms ``soft excitation process'' and ``hard scattering process'' used in this work are relative and descriptive rather than strict classifications, consistent with common practice in the field~\cite{74e,74f}. While soft excitation corresponds to small momentum transfer squared \(Q^2\) [\(<4\) (GeV/\(c\))\(^2\)] and hard scattering to large \(Q^2\) [\(>25\sim100\) (GeV/\(c\))\(^2\)], the \(Q^2\) values associated with the production of most particles in high-energy collisions are not extremely large. Using the same formalism for both processes does not introduce significant discrepancies in the extraction of key parameters. Additionally, high-energy collision processes are highly stochastic in nature, and some of the distributions we employ can be interpreted as statistical distributions rather than as strictly thermal in origin.

\section{Summary}

In summary, within the framework of multi-source thermal model, we analyze the bulk properties of multi-particle production in \(eA\) collisions at the EIC through a two-component statistical distribution. The soft excitation process involves a small number of principal contributors (sea quarks, gluons, and a lepton), while the hard scattering process involves another set of principal contributors (valence quarks or gluons, and a lepton). For both transverse momentum and multiplicity distributions, our model yields a two-component Erlang distribution. This allows us to determine the individual contributions of each principal participant and their respective numbers in the soft and hard processes.

In \(eA\) collisions, the soft excitation process characterized by lower temperatures and a \(\triangle\)-shaped topology results in net-baryons being mostly distributed in the backward rapidity region. In contrast, a Y-shaped topology in the soft process leads to net-baryons primarily distributed in the central rapidity region. For the hard scattering process, which is associated with higher temperatures, both the \(\triangle\)- and Y-shaped topologies produce net-baryons concentrated in the backward and central rapidity regions, making the two topologies indistinguishable in this case. When contributions from both soft and hard processes are combined, the \(\triangle\)-shaped topology predicts higher temperatures in the central rapidity region, while the Y-shaped topology corresponds to higher temperatures in the backward rapidity region.

The temperature trends obtained from previous comprehensive analyses of soft and hard processes align with predictions for the \(\triangle\)-shaped topology but are inconsistent with those for the Y-shaped topology. Both prior and current studies, along with the multi-source thermal model, support the valence quark-stopping scenario. This scenario can be further tested in future GeV-energy EIC experiments. At the EIC, one can measure the \(\langle p_T\rangle\) of charged particles in both backward and central rapidity regions. If \(\langle p_T\rangle\) in the central region is larger than that in the backward region, it would support the valence quark-stopping scenario; conversely, if \(\langle p_T\rangle\) in the central region is smaller, it would favor the gluon junction-stopping scenario.
\\
\\
{\bf Funding}
\\
The work of the Shanxi Group was supported by the National Natural Science Foundation of China under Grant No. 12147215, the Fundamental Research Program (Natural Science Foundation) of Shanxi Province under Grant No. 202303021212293, and the Fund for Shanxi ``1331 Project" Key Subjects Construction. The work of K.K.O. was supported by the Agency of Innovative Development under the Ministry of Higher Education, Science and Innovations of the Republic of Uzbekistan within the fundamental project No.
F3-20200929146 on analysis of open data on heavy-ion collisions at RHIC and LHC. 
\\
\\
{\bf Disclosure}
\\
The manuscript was checked for grammar and spelling using a AI
language tool, ERNIE Bot. The funding agencies have no role in the
design of the study; in the collection, analysis, or
interpretation of the data; in the writing of the manuscript; or
in the decision to publish the results. 
\\
\\
{\bf Ethical Statement}
\\
The authors declare that they are in compliance with ethical
standards regarding the content of this paper.
\\
\\
{\bf Conflicts of Interest}
\\
The authors declare no conflicts of interest.
\\
\\
{\bf Data Availability Statement}
\\
The data used to support the
findings of this study and some outcomes or conclusive statements
are included within the article and are cited at relevant places
within the text as references.
\\
\\

{\small
\begin{thebibliography}{99}

\setlength{\itemsep}{2pt}

\bibitem{1}
B. Schenke, C. Shen, and P. Tribedy, ``Bulk properties and
multi-particle correlations in large and small systems," {\it
Nuclear Physics A} 1005 (2021): 121756,
https://doi.org/10.1016/j.nuclphysa.2020.121756

\bibitem{2}
Y. C. Feng and F. Q. Wang, ``Review of nonflow estimation methods
and uncertainties in relativistic heavy-ion collisions," {\it
Journal of Physics G} 52 (2025): 013001,
https://doi.org/10.1088/1361-6471/ad903b

\bibitem{2a}
L. P. Du, A. Sorensen, and M. Stephanov, ``The QCD phase diagram
and Beam Energy Scan physics: a theory overview," {\it
International Journal of Modern Physics E} 33 (2024):
2430008,
https://doi.org/10.1142/S021830132430008X

\bibitem{2b}
T. Niida and S. A. Voloshin, ``Polarization phenomenon in
heavy-ion collisions," {\it International Journal of Modern
Physics E} 33 (2024): 2430010,
https://doi.org/10.1142/S0218301324300108

\bibitem{3}
A. Rios, A. Polls, A. Ramos, and I. Vida{\~n}a, ``Bulk and
single-particle properties of hyperonic matter at finite
temperature," {\it Physical Review C} 72 (2005): 024316,
https://doi.org/10.1103/PhysRevC.72.024316

\bibitem{3a}
S. Gupta, T. Michael, N. Bano, and A. N. Mishra, ``Study of
thermodynamic observables in Oxygen + Oxygen collisions at
$\sqrt{s_{NN}}=7$ TeV using color string percolation approach,"
{\it International Journal of Modern Physics E} 33 (2024): 2450036,
https://doi.org/10.1142/S0218301324500368

\bibitem{3b}
V. Kovalenko, ``Evolution and fluctuations of chiral chemical
potential in heavy ion collisions," {\it International Journal of
Modern Physics E} 33 (2024): 2450037,
https://doi.org/10.1142/S021830132450037X

\bibitem{4}
M. Hegazy, A. Rafaat, N. Magdy, W. L. Li, A. Deshpande, A. M. H.
Abdelhady, and A. Y. Ellithi, ``Centrality definition in e + A
collisions at the electron-ion collider," {\it Journal of Physics
G} 52 (2025): 015002,
https://doi.org/10.1088/1361-6471/ad9344

\bibitem{5}
J. F. Paquet, ``Applications of emulation and Bayesian methods in
heavy-ion physics," {\it Journal of Physics G} 51 (2024): 103001, 
https://doi.org/10.1088/1361-6471/ad6a2b

\bibitem{6}
K. J. Eskola, V. J. Kolhinen, P. V. Ruuskanen, and R. L. Thews,
``Effects of shadowing on Drell-Yan dilepton production in high
energy nuclear collisions," {\it International Journal of Modern
Physics E} 12 (2003): 197--209,
https://doi.org/10.1142/S0218301303001260

\bibitem{7}
X. Zhu, N. Xu, and P. Zhuang, ``Effect of partonic ``wind" on
charm quark correlations in high-energy nuclear collisions," {\it
Physical Review Letters} 100 (2008): 152301,
https://doi.org/10.1103/PhysRevLett.100.152301

\bibitem{8}
L. Cunqueiro, J. Dias de Deus, and C. Pajares, ``Nuclear like
effects in proton-proton collisions at high energy," {\it The
European Physical Journal C} 65 (2010): 423--426,
https://doi.org/10.1140/epjc/s10052-009-1215-8

\bibitem{9}
D. E. Kharzeev, J. Liao, S. A. Voloshin, and G. Wang, ``Chiral
magnetic and vortical effects in high-energy nuclear
collisions---A status report," {\it Progress in Particle and
Nuclear Physics} 88 (2016): 1--28,
https://doi.org/10.1016/j.ppnp.2016.01.001

\bibitem{10}
E. Basso, V. P. Goncalves, M. Krelina, J. Nemchik, and R.
Pasechnik, ``Nuclear effects in Drell-Yan pair production in
high-energy pA collisions," {\it Physical Review D} 93 (2016): 094027,
https://doi.org/10.1103/PhysRevD.93.094027

\bibitem{11}
J. M. Cornwall, ``Baryon Wilson loop area law in QCD," {\it
Physical Review D} 54 (1996): 6527--6536,
https://doi.org/10.1103/PhysRevD.54.6527

\bibitem{11a}
S. A. Bass, B. M{\"u}ller, D. K. Srivastava, ``Net baryon density
in Au+Au collisions at the Relativistic Heavy Ion Collider," {\it
Physical Review Letters} 91 (2003): 052302,
https://doi.org/10.1103/PhysRevLett.91.052302

\bibitem{11b}
Y. Mehtar-Tani and G. Wolschin, ``Baryon stopping as a new probe
of geometric scaling," {\it Physical Review Letters} 102 (2009): 182301,
https://doi.org/10.1103/PhysRevLett.102.182301

\bibitem{11c}
S. Li and S.-Q. Feng, ``Gluon saturation and baryon stopping in
the SPS, RHIC, and LHC energy regions," {\it Chinese Physics C} 36 (2012): 136--141,
https://doi.org/10.1088/1674-1137/36/2/006

\bibitem{12}
X. Artru, ``String model with baryons: Topology; classical
motion," {\it Nuclear Physics B} 85 (1975): 442--460,
https://doi.org/10.1016/0550-3213(75)90019-X

\bibitem{12a}
STAR Collaboration, ``Tracking the baryon number with nuclear collisions," 
{\it Science} 393 (2026): 727--731, 
https://www.science.org/doi/10.1126/science.ads5962
https://doi.org/10.1126/science.ads5962

\bibitem{13}
D. Kharzeev, ``Can gluons trace baryon number?," {\it Physics
Letters B} 378 (1996): 238--246,
https://doi.org/10.1016/0370-2693(96)00435-2

\bibitem{13a}
S. E. Vance, M. Gyulassy, and X. N. Wang, ``Baryon number
transport via gluonic junctions," {\it Physics Letters B} 443 (1998): 45--50,
https://doi.org/10.1016/S0370-2693(98)01338-0

\bibitem{13b}
S. E. Vance, M. Gyulassy, and X. N. Wang, ``Baryon junction
stopping at the SPS and RHIC via HIJING/B," {\it Nuclear Physics
A} 638 (1998): 395c--398c, 
https://doi.org/10.1016/S0375-9474(98)00394-7

\bibitem{14}
G. C. Rossi and G. Veneziano, ``A possible description of baryon
dynamics in dual and gauge theories," {\it Nuclear Physics B} 123 (1977): 507--545,
https://doi.org/10.1016/0550-3213(77)90178-X

\bibitem{15}
T. T. Takahashi, H. Matsufuru, Y. Nemoto, and H. Suganuma,
``Three-quark potential in SU(3) lattice QCD," {\it Physical
Review Letters} 86 (2001): 18--21,
https://doi.org/10.1103/PhysRevLett.86.18

\bibitem{16}
H. Suganuma, T. T. Takahashi, F. Okiharu, and H. Ichie, ``Lattice
QCD study for the interquark force in three-quark and multi-quark
systems," {\it AIP Conference Proceedings} 756 (2005): 123--132,
https://doi.org/10.1063/1.1920939

\bibitem{16aa}
F. Bissey, F.-G. Cao, A. R. Kitson, A. I. Signal, D. B. Leinweber,
B. G. Lasscock, and A. G. Williams, ``Gluon flux-tube distribution
and linear confinement in baryons," {\it Physical Review D} 76 (2007): 114512,
https://doi.org/10.1103/PhysRevD.76.114512

\bibitem{16a}
G. Pihan, A. Monnai, B. Schenke, and C. Shen, ``Unveiling baryon
charge carriers through charge stopping in isobar collisions,"
{\it Physical Review Letters} 133 (2024): 182301,
https://doi.org/10.1103/PhysRevLett.133.182301

\bibitem{17}
W. D. Lv, Y. Li, Z. Y. Li, R. R. Ma, Z. B. Tang, P. Tribedy, C. Y.
Tsang, Z. B. Xu, and W. M. Zha, ``Correlations of baryon and
charge stopping in heavy ion collisions," {\it Chinese Physics C} 48 (2024): 044001,
https://doi.org/10.1088/1674-1137/ad243f

\bibitem{18}
N. Magdy, A. Deshpande, R. Lacey, W. L. Li, P. Tribedy, and Z. B.
Xu, ``Search for baryon junctions in e+A collisions at the
electron ion collider," {\it The European Physical Journal C} 84 (2024): 1326,
https://doi.org/10.1140/epjc/s10052-024-13702-9

\bibitem{18a}
N. Lewis, W. D. Lv, M. A. Ross, C. Y. Tsang, J. D. Brandenburg,
Z.-W. Lin, R. R. Ma, Z. B. Tang, P. Tribedy, and Z. B. Xu,
``Search for baryon junctions in photonuclear processes and isobar
collisions at RHIC," {\it The European Physical Journal C} 84 (2024): 590,
https://doi.org/10.1140/epjc/s10052-024-12834-2

\bibitem{19}
R. L. Glauber, ``High-energy collision theory," in {\it Lectures
of Theoretical Physics}, W. E. Brittin and L. G. Dunham, Eds.,
vol. 1, pp. 315--414, Interscience Press, New York, NY, USA, 1959.

\bibitem{20}
W. Czy{\.z} and L. C. Maximon, ``High energy, small angle elastic
scattering of strongly interacting composite particles," {\it
Annals of Physics (New York)} 52 (1969): 59--121,
https://doi.org/10.1016/0003-4916(69)90321-2

\bibitem{21}
A. Bia{\l}as, M. B{\l}eszy{\'n}ski, and W. Czy{\.z}, ``Relation
between the Glauber model and classical probability calculus,"
{\it Acta Physica Polonica B} 8 (1977): 389--392,
https://www.actaphys.uj.edu.pl/fulltext?series=Reg$\&$vol=8$\&$page=389

\bibitem{22}
M. V. Ricciardi, T. Enqvist, J. Pereira, J. Benlliure, M. Bernas,
E. Casarejos, V. Henzl, A. Keli{\'c}, J. Ta{\"i}eb, and K. H.
Schmidt, ``Experimental indications for the response of the
spectators to the participant blast," {\it Physical Review
Letters} 90 (2003): 212302,
https://doi.org/10.1103/PhysRevLett.90.212302

\bibitem{23}
L. Shi, P. Danielewicz, and R. Lacey, ``Spectator response to the
participant blast," {\it Physical Review C} 64 (2001):034601,
https://doi.org/10.1103/PhysRevC.64.034601

\bibitem{24}
T. Gaitanos, H. H. Wolter, and C. Fuchs, ``Spectator and
participant decay in heavy ion collisions," {\it Physics Letters
B} 478 (2000): 79--85,
https://doi.org/10.1016/S0370-2693(00)00300-2

\bibitem{25}
M. L. Miller, K. Reygers, S. J. Sanders, and P. Steinberg,
``Glauber modeling in high-energy nuclear collisions," {\it Annual
Review of Nuclear and Particle Science} 57 (2007): 205--243,
https://doi.org/10.1146/annurev.nucl.57.090506.123020

\bibitem{26}
V. Vovchenko, D. Anchishkin, and L. P. Csernai, ``Time dependence
of partition into spectators and participants in relativistic
heavy-ion collisions," {\it Physical Review C} 90 (2014): 044907,
https://doi.org/10.1103/PhysRevC.90.044907

\bibitem{27}
A. D. Sood and R. K. Puri, ``The study of participant-spectator
matter and collision dynamics in heavy-ion collisions," {\it
International Journal of Modern Physics E} 15 (2006): 899--910,
https://doi.org/10.1142/S0218301306004685

\bibitem{28}
F. H. Liu, ``Unified description of multiplicity distributions of
final-state particles produced in collisions at high energies,"
{\it Nuclear Physics A} 810 (2008): 159--172,
https://doi.org/10.1016/j.nuclphysa.2008.06.014

\bibitem{29}
F. H. Liu, Y. Q. Gao, T. Tian, and B. C. Li, ``Unified description
of transverse momentum spectrums contributed by soft and hard
processes in high-energy nuclear collisions," {\it The European
Physical Journal A} 50 (2014): 94,
https://doi.org/10.1140/epja/i2014-14094-9

\bibitem{30}
J. Cleymans and D. Worku, ``Relativistic thermodynamics:
Transverse momentum distributions in high-energy physics," {\it
The European Physical Journal A} 48 (2012): 160,
https://doi.org/10.1140/epja/i2012-12160-0

\bibitem{31}
A. De Falco (for the ALICE Collaboration), ``Vector meson
production in pp collisions at $\sqrt{s}=7$ TeV, measured with the
ALICE detector," {\it Journal of Physics G} 38 (2011): 124083,
https://doi.org/10.1088/0954-3899/38/12/124083

\bibitem{31a}
L. D. Landau, ``Multiple production of particles under collision
of rapid particles," {\it Izvestiya Akademii Nauk: Series
Fizicheskikh}, vol. 17, pp. 51--67, 1953 (in Russian), in {\it
English-Translation: Collected Papers of L. D. Landau}, D.
Ter-Haarp, Ed., p. 569, Pergamon Press, Oxford, UK, 1965.

\bibitem{31b}
P. A. Steinberg, ``Bulk dynamics in heavy ion collisions," {\it
Nuclear Physics A} 752 (2005): 423--432,
https://doi.org/10.1016/j.nuclphysa.2005.02.139

\bibitem{31c}
L. N. Gao and F. H. Liu, ``On pseudorapidity distribution and
speed of sound in high energy heavy ion collisions based on a new
revised Landau hydrodynamic model," {\it Advances in High Energy
Physics}, vol. 2015, Article ID 184713, 23 pages, 2015.
https://doi.org/10.1155/2015/184713

\bibitem{31d}
L. N. Gao and F. H. Liu, ``On distributions of emission sources
and speed-of-sound in proton-proton (proton-antiproton)
collisions," {\it Advances in High Energy Physics} 2015 (2015): 641906,
https://doi.org/10.1155/2015/641906

\bibitem{31e}
C.-Y. Wong, ``Landau hydrodynamics reexamined," {\it Physical
Review C} 78 (2008): 054902,
https://doi.org/10.1103/PhysRevC.78.054902

\bibitem{32}
E. K. G. Sarkisyan and A. S. Sakharov, ``Multihadron production
features in different reactions," {\it AIP Conference
Proceedings} 828 (2006): 35--41,
https://doi.org/10.1063/1.2197392

\bibitem{33}
E. K. G. Sarkisyan and A. S. Sakharov, ``Relating multihadron
production in hadronic and nuclear collisions," {\it The European
Physical Journal C} 70 (2010): 533--541,
https://doi.org/10.1140/epjc/s10052-010-1493-1

\bibitem{34}
E. K. G. Sarkisyan, A. N. Mishra, R. Sahoo, and A. S. Sakharov,
``Multihadron production dynamics exploring the energy balance in
hadronic and nuclear collisions," {\it Physical Review D} 93 (2016): 054046,
https://doi.org/10.1103/PhysRevD.93.054046

\bibitem{35}
A. N. Mishra, R. Sahoo, E. K. G. Sarkisyan, and A. S. Sakharov,
``Effective-energy budget in multiparticle production in nuclear
collisions," {\it The European Physical Journal C} 74 (2014): 3147,
https://doi.org/10.1140/epjc/s10052-014-3147-1

\bibitem{36}
E. K. G. Sarkisyan, A. N. Mishra, R. Sahoo, and A. S. Sakharov,
``Centrality dependence of midrapidity density from GeV to TeV
heavy-ion collisions in the effective-energy universality picture
of hadroproduction," {\it Physical Review D} 94 (2016): 011501,
https://doi.org/10.1103/PhysRevD.94.011501

\bibitem{37}
E. K. Sarkisyan-Grinbaum, A. N. Mishra, R. Sahoo, and A. S.
Sakharov, ``Effective-energy universality approach describing
total multiplicity centrality dependence in heavy-ion collisions,"
{\it EPL} 127 (2019): 62001,
https://doi.org/10.1209/0295-5075/127/62001

\bibitem{38}
A. N. Mishra, A. Ortiz, and G. Pai{\'c}, ``Intriguing similarities
of high-$p_T$ particle production between $pp$ and $A$-$A$
collisions,"{\it Physical Review C} 99 (2019): 034911,
https://doi.org/10.1103/PhysRevC.99.034911

\bibitem{39}
P. Castorina, A. Iorio, D. Lanteri, H. Satz, and M. Spousta,
``Universality in hadronic and nuclear collisions at high energy,"
{\it Physical Review C} 101 (2020): 054902,
https://doi.org/10.1103/PhysRevC.101.054902

\bibitem{39a}
P. Braun-Munzinger, J. Stachel, J. P. Wessels, and N. Xu,
``Thermal equilibration and expansion in nucleus-nucleus
collisions at the AGS," {\it Physics Letters B} 344 (1995): 43--48,
https://doi.org/10.1016/0370-2693(94)01534-J

\bibitem{39b}
A. Andronic, P. Braun-Munzinger, and J. Stachel, ``Thermal hadron
production in relativistic nuclear collisions: The hadron mass
spectrum, the horn, and the QCD phase transition," {\it Physics
Letters B} 673 (2009): 142--145,
https://doi.org/10.1016/j.physletb.2009.02.014

\bibitem{39c}
STAR Collab. (B. I. Abelev {\it et al.}), ``Systematic
measurements of identified particle spectra in pp, d+Au, and Au+Au
collisions at the STAR detector," {\it Physical Review C} 79 (2009): 034909,
https://doi.org/10.1103/PhysRevC.79.034909

\bibitem{39d}
J. Cleymans, H. Oeschler, and K. Redlich, ``Influence of impact
parameter on thermal description of relativistic heavy ion
collisions at (1--2)$A$ GeV," {\it Physical Review C} 59 (1999): 1663--1673,
https://doi.org/10.1103/PhysRevC.59.1663

\bibitem{39e}
P. Braun-Munzinger, I. Heppe, and J. Stachel, ``Chemical
equilibration in Pb+Pb collisions at the SPS," {\it Physics
Letters B} 465 (1999): 15--20,
https://doi.org/10.1016/S0370-2693(99)01076-X

\bibitem{39f}
J. Manninen and F. Becattini, ``Chemical freeze-out in
ultrarelativistic heavy ion collisions at $\sqrt{s_{NN}}=130$ and
200 GeV," {\it Physical Review C} 78 (2008): 054901,
https://doi.org/10.1103/PhysRevC.78.054901

\bibitem{39g}
A. Andronic, P. Braun-Munzinger, and K. Redlich, ``Decoding the
phase structure of QCD via particle production at high energy,"
{\it Nature} 561 (2018): 321--330,
https://doi.org/10.1038/s41586-018-0491-6

\bibitem{39h}
PHENIX Collab. (Adler S. S. {\it et al.}), ``Identified charged
particle spectra and yields in Au+Au collisions at
$\sqrt{s_{NN}}=200$ GeV," {\it Physical Review C} 69 (2004): 034909,
https://doi.org/10.1103/PhysRevC.69.034909

\bibitem{39i}
P. Koch, J. Rafelski, and W. Greiner, ``Strange hadrons in hot
nuclear matter," {\it Physics Letters B} 123 (1983): 151--154,
https://doi.org/10.1016/0370-2693(83)90411-2

\bibitem{39j}
P. Braun-Munzinger, D. Magestro, K. Redlich, and J. Stachel,
``Hadron production in Au-Au collisions at RHIC," {\it Physics
Letters B} 518 (2001): 41--46,
https://doi.org/10.1016/S0370-2693(01)01069-3

\bibitem{39k}
H. L. Lao, Y. Q. Gao, and F. H. Liu, ``Energy dependent chemical
potentials of light particles and quarks from yield ratios of
antiparticles to particles in high energy collisions," {\it
Universe} 5 (2019): 152,
https://doi.org/10.3390/universe5060152

\bibitem{39l}
H. L. Lao, Y. Q. Gao, and F. H. Liu, ``Light particle and quark
chemical potentials from negatively to positively charged particle
yield ratios corrected by removing strong and weak decays," {\it
Advances in High Energy Physics} 2020 (2020): 5064737,
https://doi.org/10.1155/2020/5064737

\bibitem{39m}
NA49 Collab. (H. Appelsh{\"a}user {\it et al.}), ``Baryon stopping
and charged particle distributions in central Pb+Pb collisions at
158 GeV per nucleon," {\it Physical Review Letters} 82 (1999): 
2471--2475,
https://doi.org/10.1103/PhysRevLett.82.2471

\bibitem{39n}
BRAHMS Collab. (I.C. Arsene {\it et al.}), ``Nuclear stopping and
rapidity loss in Au+Au collisions at $\sqrt{s_{NN}}=62.4$ GeV,"
{\it Physics Letters B} 677 (2009): 267--271,
https://doi.org/10.1016/j.physletb.2009.05.049

\bibitem{39o}
F. Videb$\ae$k (for the BRAHMS Collab.), ``Overview and recent
results from BRAHMS," {\it Nuclear Physics A} 830 (2009): 
43c--50c,
https://doi.org/10.1016/j.nuclphysa.2009.09.011

\bibitem{39p}
BRAHMS Collab. (I. G. Bearden {\it et al.}), ``Nuclear stopping in
Au+Au collisions at $\sqrt{s_{NN}}=200$ GeV," {\it Physical Review
Letters} 93 (2004): 102301,
https://doi.org/10.1103/PhysRevLett.93.102301

\bibitem{40}
Y. H. Chen, F. H. Liu, and E. K. Sarkisyan-Grinbaum, ``Event
patterns from negative pion spectra in proton-proton and
nucleus-nucleus collisions at SPS," {\it Chinine Physics C} 42 (2018): 104102,
https://doi.org/10.1088/1674-1137/42/10/104102

\bibitem{41}
P. P. Yang, F. H. Liu, and K. K. Olimov, ``Rapidity and energy
dependencies of temperatures and volume extracted from identified
charged hadron spectra in proton-proton collisions at a super
proton synchrotron (SPS)," {\it Entropy} 25 (2023): 1571,
https://doi.org/10.3390/e25121571

\bibitem{42}
P. P. Yang, M. Ajaz, M. Waqas, F. H. Liu, and M. K. Suleymanov,
``Pseudorapidity dependence of the $p_T$ spectra of charged
hadrons in pp collisions at $\sqrt{s}=0.9$ and 2.36 TeV," {\it
Journal of Physics G} 49 (2022): 055110,
https://doi.org/10.1088/1361-6471/ac5d0b

\bibitem{43}
M. Waqas, M. Ajaz, A. H. Ismail, A. Tawfik, M. B. Ammar, and H. I.
Alrebdi, ``Bulk properties of charged particles as a function of
pseudo-rapidity in pp collisions," {\it The European Physical
Journal A} 60 (2024): 123,
https://doi.org/10.1140/epja/s10050-024-01338-y

\bibitem{73}
F. H. Liu, ``Transverse-energy distribution in proton-nucleus
collisions at high energy," {\it Canadian Journal of Physics} 79 (2001): 739--748,
https://doi.org/10.1139/cjp-79-4-739

\bibitem{73a}
C. Alexandrou, S. Bacchio, M. Constantinou, J. Finkenrath, K.
Hadjiyiannakou, K. Jansen, G. Koutsou, H. Panagopoulos, and G.
Spanoudes, ``Complete flavor decomposition of the spin and
momentum fraction of the proton using lattice QCD simulations at
physical pion mass," {\it Physical Review D} 101 (2020): 094513,
https://doi.org/10.1103/PhysRevD.101.094513

\bibitem{73b}
P. C. Tandy, ``Parton decomposition of nucleon spin and momentum:
Gluons from dressed quarks," {\it Physics Letters B} 842 (2023): 137972,
https://doi.org/10.1016/j.physletb.2023.137972

\bibitem{74}
ALICE Collab. (S. Acharya {\it et al.}), ``Measurement of
correlations among net-charge, net-proton, and net-kaon
multiplicity distributions in Pb-Pb collisions at
$\sqrt{s_{NN}}=5.02$ TeV," {\it Journal of High Energy Physics} 2025(08) (2025): 210,
https://doi.org/10.1007/JHEP08(2025)210

\bibitem{74a}
S. A. Bass, M. Belkacem, M. Bleicher, M. Brandstetter, L. Bravina,
C. Ernst, L. Gerland, M. Hofmann, S. Hofmann, J. Konopka, G. Mao,
L. Neise, S. Soff, C. Spieles, H. Weber, L. A. Winckelmann, H.
St{\"o}cker, W. Greiner, Ch. Hartnack, J. Aichelin, and N.
Amelin, ``Microscopic models for ultrarelativistic heavy ion
collisions," {\it Progress in Particle and Nuclear Physics} 41 (1998): 255--369,
https://doi.org/10.1016/S0146-6410(98)00058-1

\bibitem{74b}
H. Petersen, J. Steinheimer, G. Burau, M. Bleicher, and H.
St{\"o}cker, ``Fully integrated transport approach to heavy ion
reactions with an intermediate hydrodynamic stage," {\it Physical
Review C} 78 (2008): 044901,
https://doi.org/10.1103/PhysRevC.78.044901

\bibitem{74c}
M. Bleicher, E. Zabrodin, C. Spieles, S. A. Bass, C. Ernst, S.
Soff, L. Bravina, M. Belkacem, H. Weber, H. St{\"o}cker, and W
Greiner, ``Relativistic hadron-hadron collisions in the
ultra-relativistic quantum molecular dynamics model," {\it Journal
of Physics G} 25 (1999): 1859--1896,
https://doi.org/10.1088/0954-3899/25/9/308

\bibitem{74d}
K. G. Boreskov, ``Probabilistic model of Reggeon Field Theory," in
{\it Multiple Facets of Quantization and Supersymmetry / Michael
Marinov Memorial Volume}, World Scientific, Singapore, pp.
322--351, 2002,
https://doi.org/10.1142/9789812777065\_0025

\bibitem{74e}
STAR Collab. (M. I. Abdulhamid {\it et al.}), ``Correlations of
event activity with hard and soft processes in p+Au collisions at
$\sqrt{s_{NN}}=200$ GeV at STAR," {\it Physical Review C} 110 (2024): 044908,
https://doi.org/10.1103/PhysRevC.110.044908

\bibitem{74f}
V. A. Bednyakov, A. A. Grinyuk, G. I. Lykasov, and M. Poghosyan,
``Role of gluons in soft and semi-hard multiple hadron production
in pp collisions at LHC," {\it International Journal of Modern
Physics A} 27 (2012): 1250042,
https://doi.org/10.1142/S0217751X1250042X


\end{thebibliography}}
\end{document}